\documentstyle[psfig]{article}
\def\begeq{\begin{equation}}
\def\endeq{\end{equation}}

\def\begeqar{\begin{eqnarray}}
\def\endeqar{\end{eqnarray}}

\def\t{\theta}
\def\a{\alpha}

\begin{document}
\begin{titlepage}
\title{Lectures on Non Perturbative Field Theory\\
 and Quantum Impurity Problems: Part II.}
\author{H. Saleur\thanks{
Department of Physics and Caltech-USC Center for Theoretical Physics, University of Southern California, 
Los-Angeles, CA 90089-0484, USA. email: saleur@physics.usc.edu}}
\date{\today}
\maketitle

\begin{abstract}
These are notes of lectures given at The NATO Advanced Study Institute/EC
Summer School on ``New Theoretical Approaches to Strongly Correlated Systems'',
(Newton Institute, April 2000). 
They are a sequel to the notes I wrote two years ago 
for the  Summer School, ``Topological Aspects of Low Dimensional
Systems'', (Les Houches, July 1998). In  this second part, I 
review the form-factors technique
and its extension to massless quantum field theories. I 
then discuss 
the calculation of correlators in integrable quantum impurity problems,
with special emphasis on point contact tunneling in the fractional
quantum Hall effect, and the two-state problem of dissipative quantum 
mechanics. 
 
\end{abstract}

\end{titlepage}

\huge \noindent{\bf Introduction}
\normalsize
\bigskip
\bigskip

I pick up the discussion of quantum impurity problems where it is left in 
my Lecture Notes form the 1998 Les Houches Summer School \cite{Houches98} 
 - consult 
these notes for 
the necessary physical 
and technical background. I have so far  explained how  integrability provides a convenient basis of quasiparticles which scatter nicely
among one another and at the impurity. I have showed how to  use this basis 
to compute DC transport properties with a combination of thermodynamic Bethe ansatz
and Landauer type arguments.
To proceed, I now would like to discuss correlation functions.

It is a natural idea
to compute correlators using the  same basis of quasiparticles,
and inserting completude relations in the Green functions of interest. 
In this 
approach, the key ingredient is the matrix elements of the physical operators. Despite
the ``almost free'' structure of the DC transport properties, 
a rather unpleasant surprise is met here however: these matrix elements are all non zero in general,
that is, physical operators (eg the current) are able to create/destroy  arbitrarily large numbers
of particles  when they act on an arbitrary state. 

These matrix  elements - usually called form-factors - turn out to be a major topic 
\cite{Karowski78,Smirnov,Mussardo94} in  recent developments (for even more 
recent works, see \cite{Others}) 
about integrable quantum field theories, and are known for a large variety of integrable
masssive quantum field theories. 

The idea to compute correlators in integrable quantum impurity problems is first to 
extend the form-factors approach to the case of theories which are massless in the bulk. The next step
is to take the impurity (boundary) interaction into account properly, and finally, to sum
over all the intermediate states to obtain physical correlations. The last step is the most difficult,
and cannot be done analytically for the moment. Fortunately, and somewhat surprisingly,
it turns out that the sums over intermediate states converge very quickly, everywhere along the 
trajectory between the UV and IR fixed points. As a result, the form-factors approach gives extremely
accurate results (and with a controlled accuracy) for most correlators of interest. 

In some cases, the form-factors expansion is plagued with an ``infrared catastrophy''
due to the proliferation of soft modes in the 
 massless scattering description. This problem
can usually be controlled by simple means, and meaningful correlators still obtained.

The main caveat of the form-factors approach is that it seems badly behaved when one approaches
isotropic limits: as a result, although correlators for the anisotropic Kondo model
up to $g=3/4$ or so are now determined, the Kondo limit is still unaccessible. This is
not much of a problem in applications to the fractional quantum Hall effect, where 
the interesting filling fractions are far from the isotropic limit of the underlying field theories. 

\section{Some generalities on form-factors}

I follow here the notations and conventions of \cite{Houches98}, in particular section 5 of the latter notes.

The space of states is generated by the vectors
\begeq
\left|\a_1,\ldots,\a_n\right>_{a_1,...,a_n}= Z^\dagger_{a_1}(\a_1)\ldots
Z^\dagger_{a_n}(\a_n)|0>
\endeq
and the dual space by
\begeq
^{a_n,...,a_1}\left<\a_n,\ldots,\a_1\right|= \left<0\right| Z^{a_n}(\a_n)\ldots
Z^{a_1}(\a_1)
\endeq
Here, the labels $a$ stand for instance for solitons/antisolitons ($\pm$) and breathers 
in the sine-Gordon model, $\alpha$ denotes the rapidity, and the $Z,Z^\dagger$ are the annihilation creation operators of the Faddeev Zamolodchikov algebra \cite{ZamoZamo79} . For a given set of rapidities and quantum numbers, the states with different orderings are related
through S matrix elements. One usually call ``in'' states the states for which $\a_1>\a_2\ldots>\a_n$,
and ``out'' states those for which $\a_1<\a_2\ldots<\a_n$. In or out states form a complete basis of the set of states, but it 
is convenient to keep some redundancy and consider all possible rapidity orderings, after division 
by the appropriate degeneracy factors.

 Form-factors
(more correctly,  ``generalized form-factors'' since no order of the rapidities
is prescribed) of an operator ${\cal O}$  in a bulk theory are
defined as~:
\begeq
f(\a_1,...,\a_n)_{a_1,...,a_n}=
<0|{\cal O}(0,0)|Z^\dagger_{a_1}(\a_1)\ldots
Z^\dagger_{a_n}(\a_n)|0>\label{ffacteg}
\endeq
where $|0>$ is the ground state. We chose to take the operator at the origin,
since relativistic invariance determines trivially the dependence of the matrix elements on the coordinates of insertion (we set $\hbar=1$)
\begeq
<0|{\cal O}(x,t)| Z^\dagger_{a_1}(\a_1)\ldots
Z^\dagger_{a_n}(\a_n)|0>= e^{i(Px-Et)} f(\a_1,...,\a_n)_{a_1,...,a_n}\label{ffactegi}
\endeq

$S$ matrices have in many cases been obtained by explicit solution of properly regularized
quantum field theories. The strategy, which was implemented quite succesfully in the 
massive Thirring model for instance \cite{Korepin}, is to find out eigenstates, fill the ground state, 
determine the possible excitations over it, and finally compute their scattering,  which is the 
result of a combination of the bare scattering and ``dressing'' effects coming from interaction 
with the ground state. Needless to say, the procedure is laborious, 
and it has proven much faster to obtain the $S$ matrices in a more abstract way, 
using general axioms of $S$ matrix theory, factorizability, and assumptions of maximal analyticity \cite{ZamoZamo79,Mussardo92,Dorey98}. 
  
It is even harder to obtain the matrix elements from the solution of bare theories. A program to do this 
is under way \cite{Korepin,Kyoto}, but very few results of interest for our problem 
have been obtained so far. Form-factors can however be determined 
(though not quite as easily as $S$ matrices) using an axiomatic approach similar in 
spirit to what was done for the $S$ matrices, and this is what we would like to discuss briefly here. 

\bigskip

One of the consequences of integrability is that multiple particle processes 
can be reexpressed in terms of two-particle ones. The basic objects in this approach are therefore
 those involving only a pair of particles. To start, consider 
the two particle $S$ matrix: by relativistic invariance, it is expected to depend only on 
the Mandelstam variable 
\begeq
s=(p_a(\a_1)+p_b(\a_2))^2=M_1^2+M_2^2+2M_1M_2\cosh(\a_1-\a_2)
\endeq
The function $S(s)$ is expected to be an  analytic function of $s$, with a cut along the real
axis running from $s=(M_1+M_2)^2$ to $s=\infty$, and another running from $s=-\infty$ to $s=(M_1-M_2)^2$. The exchange of in and out states
corresponds to exchanging the upper lip of the right cut $\hbox{Im }s=0+$ with the 
lower lip of the right cut $\hbox{Im }s=0-$. Crossing instead exchanges the upper lip of the 
right cut with the lower lip of the left cut. 

This plane with the 
cuts transforms into the ``physical'' strip $0\leq\hbox{Im }\a\leq\pi$ in terms of the variable $\a=\a_1-\a_2$. The line $\hbox{Im }\a=0$ 
corresponds to the right cut, while $\hbox{Im }\a=\pi$ corresponds to the left cut. Upper and lower lips correspond 
to $\hbox{Re }\a>0$, resp. $\hbox{Re }\a<0$. 

In terms of the $\a$ variable, the standard properties of the $S$ matrix are: 
\begeqar
&\bullet& S_{a_1a_2}^{a'_1a'_2}(\a)=S_{a_2a_1}^{a'_2a'_1}(\a)\ \ \ \ \hbox{C}\nonumber\\
&\bullet& S_{a_1a_2}^{a'_1a'_2}(\a)=S_{\bar{a}_1\bar{a}_2}^{\bar{a}'_1\bar{a}'_2}(\a)\ \ \ \ \hbox{P}\nonumber\\
&\bullet& S_{a_1a_2}^{a'_1a'_2}(\a)=S_{a'_2a'_1}^{a_2a_1}(\a)\ \ \ \ \hbox{T}
\endeqar
and
\begeqar
&\bullet& \hbox{$S(\a)$ is real for $\a$ purely imaginary}\nonumber\\
&\bullet& \hbox{Unitarity}\ S_{a_1a_2}^{b_1b_2}(\a)S_{b_1b_2}^{c_1c_2}(-\a)=\delta_{a_1}^{c_1}\delta_{a_2}^{c_2}\nonumber\\
&\bullet& \hbox{Crossing}\ S_{a_1a_2}^{b_1b_2}(\a)=S_{a_1\bar{b}_2}^{b_1\bar{a}_2}(i\pi-\a)
\endeqar
In addition of course, the $S$ matrix is a solution of the Yang Baxter
equation, which reads in components
\begeq
S_{a_1a_2}^{b_1b_2}(\a_1-\a_2)S_{b_1a_3}^{c_1b_3}(\a_1-\a_3)
S_{b_2b_3}^{c_2c_3}(\a_2-\a_3)=S_{a_2a_3}^{b_2b_3}(\a_2-\a_3)
S_{a_1b_3}^{b_1c_3}(\a_1-\a_3)S_{b_1b_2}^{c_1c_2}(\a_1-\a_2)
\endeq

The space of states is conveniently built using 
 the Faddeev Zamolodchikov algebra
\begeqar
Z^{a_1}(\a_1)Z^{a_2}(\a_2)&=&S^{a_1a_2}_{a_1'a_2'}(\a_1-\a_2)Z^{a'_2}
(\a_2)Z^{a'_1}(\a_1)\nonumber\\
Z^\dagger_{a_1}(\a_1)Z^\dagger_{a_2}(\a_2)&=&S_{a_1a_2}^{a'_1a'_2}(\a_1-\a_2)
Z^\dagger_{a'_2}
(\a_2)Z^\dagger_{a'_1}(\a_1)\nonumber\\
Z^{a_1}(\a_1)Z_{a_2}^\dagger(\a_2)&=&S_{a_2a'_1}^{a'_2a_1}(\a_1-\a_2)Z^\dagger_{a'_2}
(\a_2)Z^{a'_1}(\a_1)+2\pi\delta^{a_1}_{a_2}\delta(\a_1-\a_2)\label{zfdef}
\endeqar
for which several of the $S$ matrix properties  (but not 
those involving crossing) are just consistency relations.

In terms of $\a$, $S$ is a meromorphic function, whose only singularities in the physical strip 
are poles on the imaginary axis, corresponding in general to bound states.  
\bigskip

Consider now the two particle form-factor $f(\a_1,\a_2)_{a_1,a_2}$. Up to some simple dimensionful terms (which are absent if the operator is scalar), this is also a function
of the Mandelstam variable $s$.  Its analytical properties 
are similar although a bit different than for $S$; in particular, the left cut does not 
exist for form-factors. 

From the second relation of (\ref{zfdef}), it follows immediately that
\begeq
f(\a_1,\a_2)_{a_1,a_2}=S_{a'_1a'_2}^{a_1a_2}(\a_1-\a_2)
f(\a_2,\a_1)_{a'_2,a'_1}\label{twoparti}
\endeq
If $\a_1>\a_2$, the state $|\a_1,\a_2>_{a_1,a_2}$ is an in state, with the corresponding out state $|\a_2,\a_1>_{a_2,a_1}$.
The exchange of the two on the other hand corresponds to going from the upper to the lower lip
of the cut in the s-plane, ie from $\hbox{Im }\a=0$ to $\hbox{Im }\a=2\pi$ in the $\a$ plane. It follows that
\begeq
f(\a_1+2i\pi,\a_2)_{a_1,a_2}=f(\a_2,\a_1)_{a_2,a_1}\label{twopartii}
\endeq

Equations (\ref{twoparti},\ref{twopartii}) are the basic tool 
to determine form-factors. To see this, suppose  for simplicity that the two particle S matrix is diagonal (actually, the following would also hold
if it could be diagonalized
by a rapidity independent change of basis: such is the case for the sine-Gordon model in the 
$+-$ sector). This would correspond, for instance, to the sinh-Gordon case, which we discuss in details below. 
Assuming the operator of interest is scalar,  $f(\a)$ is
thus found to satisfy the constraints
\begeqar
f(\a)=S(\a)f(-\a)\nonumber\\
f(2i\pi+\a)=f(-\a)\label{basicffeq}
\endeqar
The minimal solution of these equations is obtained by assuming that $f$ does
not have poles nor zeroes (except the threshold pole  at $\a=0$) in the 
strip $\hbox{Im } \a\in [0,2\pi]$. Consider 
now a closed contour ${\cal C}$ enclosing this strip. By Cauchy's theorem
one has
\begeqar
{d\over d\a} \ln f^{min}(\a)&=&{1\over 8i\pi}\int_{{\cal C}} {dz\over \sinh^2 (z-\a)/2}\ln f(z)\nonumber\\
&=& {1\over 8i\pi}\int_{-\infty}^\infty  {dz\over \sinh^2 (z-\a)/2}\ln {f(z)\over f(z+2i\pi)}\nonumber\\
&=& {1\over 8i\pi}\int_{-\infty}^\infty  {dz\over \sinh^2 (z-\a)/2}\ln S(z)
\endeqar
Very often, the S matrix can be recast  in the form
\begeq
S(\a)=-\exp \left[\int_0^\infty dx g(x) \sinh (x\a/ i\pi)\right]\label{shape}
\endeq
from which one finds
\begeq
f^{min}=\hbox{cst }\sinh{\a\over 2}\exp\left\{
 \int_0^\infty dx {g(x)\over\sinh x} \sin^2[ x(i\pi-\a)/2\pi]\right\}
\endeq
In general, form-factors are expressed in terms of $f^{min}$ and simple 
functions which encode the required pole structure, see below.

Generalizing the two equations (\ref{basicffeq}) we have, for $n$ particles,
\begeqar
f(\a_1,\ldots,\a_i,\a_{i+1},\ldots,\a_n)_{a_1,\ldots,a_i,a_{i+1},\ldots,a_n}&=&f(\a_1,\ldots,\a_{i+1},\a_i,\ldots,\a_n)_{a_1,\ldots,b_{i+1},b_i,\ldots,a_n}\nonumber\\
&\times& 
S_{a_ia_{i+1}}^{b_ib_{i+1}}(\a_i-\a_{i+1})\nonumber\\
f_{a_1\ldots a_n}(\a_1+2i\pi,\ldots,\a_n)&=&f_{a_2\ldots a_n,a_1}(\a_2,\ldots,\a_n,\a_1)
\label{genffeq}
\endeqar

The form-factors are meromorphic functions of the rapidity differences $\a_{ij}$ in the strip $0\leq \hbox{Im }\a_{ij}\leq 2\pi$, with two kinds of simple poles. 

One type of poles is called annihilation, or kinematical, pole. Such 
poles are always expected, even if the theory has no bound states, 
 when some of the rapidities in the form-factor  differ by $i\pi$, corresponding
physically to the presence of a pair particle-antiparticle in the process. The residue is a form-factor with two fewer particles
\begeq
i\hbox{ Res}_{\a'=\a+i\pi}
f(\a',\a,\a_1,\a_2,\ldots,\a_n)_{\bar{a}aa_1\ldots a_n}=
\left[\delta_{a_1}^{b_1}\ldots\delta_{a_n}^{b_n}-
S_{a_1\ldots a_n}^{b_1\ldots b_n}(\a_1\ldots \a_n|\a)\right]
f(\a_1,\ldots,\a_n)_{b_1,\ldots, b_n},\label{residuei}
\endeq
where, we have defined the $S$ matrix element
\begeq
S_{a_1\ldots a_n}^{b_1\ldots b_n}(\a_1\ldots \a_n|\a)=
S_{a_1c_n}^{b_1c_1}(\a_1-\a)S_{a_2c_1}^{b_2c_2}(\a_2-\a)\ldots
S_{a_nc_{n-1}}^{b_nc_n}(\a_n-\a)
\endeq
This provides a resursive relation between form-factors with $n+2$ and $n$ particles, which proves crucial in determining the multiple particle form-factors.

The other type of poles has a more physical origin, and corresponds to the 
appearance of bound states. In words,  form-factors obey relations that mimic the bootstrap structure of the 
theory: if a particle appears as a bound state of two others, its form-factors are residues of 
form-factors involving these particles. In formulas, things are a bit more complicated. If $c$ is a bound state of particles $a$ and $b$, such that the $S$ matrix has a simple pole
\begeq
S_{ab}^{de}(\a)\approx {i g_{ab}^c g_c^{de}\over \a-iu_{ab}^c}
\endeq
then the form factor $f(\a_1,\a_2,\ldots,\a_{n+2})_{a_1\ldots a_nab}$
has a pole when $\a_{n+1}-\a_{n+2}=iu_{ab}^c$, with residue
\begeq
i\hbox{Res }_{\a_{n+1}-\a_{n+2}=iu_{ab}^c}f(\a_1,\a_2,\ldots,\a_{n+2})_{a_1\ldots a_nab}=g_{ab}^c f(\a_1,\a_2,\ldots,\a_n,\a_{n+1}-i u_{ab}^c/2)_{a_1\ldots a_nc}\label{residueii}
\endeq
where the value of the $n+1^{th}$ rapidity 
follows from  conservation of energy and momentum
at the three particle vertex $ab\rightarrow c$. Formulas (
\ref{residuei},\ref{residueii}) are usually called LSZ reduction formulas, from 
the pioneering paper of Lehmann, Symanzik and Zimmermann \cite{LSZ}.

In the simple case where there is only one type of particle (like in 
the sinh-Gordon model, see below), the general solution of these equations can always be written in the form
\begeq
f(\a_1,\ldots,\a_n)=K(\a_1,\ldots,\a_n)\prod_{i<j} f^{min}(\a_i-\a_j)
\endeq
where $K$ is a completely symmetric, doubly periodic function, which contains all the physical poles. 

Finally, we point out that other form-factors can be obtained by crossing
\begeq
_{a'_1,\ldots, a'_m}\left<\a'_1,\ldots, \a'_m|{\cal O}|\a_1,\ldots,\a_n\right>_{a_1,\ldots,a_n}=f(\a'_1+i\pi,\ldots,a'_m+i\pi,\a_1,\ldots,\a_n)_{\bar{a}_1',\ldots,\\bar{a}_m',a_1,\ldots,a_n}
\endeq

\bigskip

Let us now discuss indeed the  sinh-Gordon model in more details - here,
I follow closely the paper \cite{Fring}. The action is
\begeq
S={1\over 16\pi g}\int_{-\infty}^\infty
dxdy\left[\left(\partial_x
\Phi\right)^2
+ \left(\partial_y\Phi\right)^2+\Lambda\cosh \Phi\right].
\label{shgoract}
\endeq
Note that here the free boson is normalized differently than in \cite{Houches98}:
this will prove more convenient below.
Eq. (\ref{shgoract}) defines an integrable massive theory; the conformal weights
of the perturbing operator are $h=\bar{h}=-g$.  The spectrum is very
simple and
consists of a  single particle of mass $M$,  and S matrix~:
\begeq
S(\alpha)={\tanh{1\over 2}\left(\alpha-i {\pi B\over
2}\right)\over
\tanh{1\over 2}\left(\alpha+i {\pi B\over 2}\right)},
\label{smat}
\endeq
where~:
$$
B={2g\over 1+g}
$$
Observe the remarkable  duality of the S matrix in $B\to 2-B$, i.e. in $g\to 1/g$. This duality is 
certainly not obvious at the level of the action, and is deeply 
 non perturbative in nature. 

To determine the minimal form-factor, it is convenient to replace trigonometric functions 
with $\Gamma$ functions using the basic identity
\begeq
\Gamma(z)\Gamma(1-z)={\pi\over \sin \pi z}
\endeq
together with the integral representation
\begeq
\ln\Gamma(z)=\int_0^\infty \left[(z-1)e^{-t}+{e^{-tz}-e^{-t}\over 1-e^{-t}}\right]{dt\over t}
\endeq
One finds then
\begeq
S(\alpha)=-\exp\left[2\int_0^\infty {dx\over x}{\cosh x(1-B)/2\over \cosh x/2}
\sinh (\a x/i\pi)\right]
\endeq
Using a similar representation for $\ln\sinh \a/2$, we find finally
\begeq
f^{min}(\a)={\cal N}\exp\left\{8\int_0^\infty {dx\over
x}
{\sinh ( {xB\over 4})\sinh [{x(2-B)\over 4}]\sinh ({x\over 2})\over
 \sinh^2(x)}\sin^2 [x(i\pi-\a)/ 2\pi]
\right\},
\endeq
Here, we have put the normalization which is standard in the literature
\cite{Fring}:
$$
{\cal N}= \exp\left[-4\int_0^\infty {dx\over x}
{\sinh ({xB\over 4})\sinh [{x(2-B)\over 4}]\sinh ({x\over 2})\over
 \sinh^2(x)}\right]
$$
Consider now the form-factors of the field $\Phi$ itself. Since $\Phi$, as well
as the creation operators of the sinh-Gordon particle, are odd under the $Z_2$ symmetry $\Phi\rightarrow -\Phi$,
only form-factors witn an even number of particles are non vanishing. One finds
the general formula \cite{Fring}
\begeq
f(\a_1,\ldots,\a_{2n+1})=\mu
\left({4\sin {\pi B\over 2} \over F^{min}(i\pi,B)}\right)^n
\sigma_{2n+1}^{(2n+1)}P_{2n+1}(x_1,\ldots,x_{2n+1})
\prod_{i<j}{f^{min}(\a_i-\a_j) \over x_i+x_j},
\label{ffexpr}
\endeq
where we introduced $x\equiv e^\a$ and the $\sigma$'s are the
basic
symmetric polynomials~:
$$
\sigma_p=\sum_{i_1<i_2<\cdots <i_p} x_{i_1} x_{i_2}\cdots x_{i_p},
$$
with the convention $\sigma_0=1$ and $\sigma_p=0$ if $p$ is greater
than the number of variables.
The $P_{2n+1}$'s are symmetric polynomials, which can be obtained
by solving LSZ \cite{LSZ} recursion relations. The first ones read~:
\begeqar
P_3(x_1,\ldots,x_3)=&1\nonumber\\
P_5(x_1,\ldots,x_5)=&\sigma_2\sigma_3-c_1^2\sigma_5\nonumber\\
P_7(x_1,\ldots,x_7)=&\sigma_2\sigma_3\sigma_4\sigma_5-
c_1^2(\sigma_4\sigma_5^2
+\sigma_1\sigma_2\sigma_5\sigma_6+
\sigma_2^2\sigma_3-c_1^2\sigma_2\sigma_5
)\nonumber\\
&-c_2(\sigma_1\sigma_6\sigma_7+\sigma_1\sigma_2\sigma_4\sigma_7)+
\sigma_3\sigma_5
\sigma_6)+c_1c_2^2\sigma_7^2.
\label{poly}
\endeqar
with $c_1=2\cos \pi B/2$, $c_2=1-c_1^2$. Observe that except for the
overall
normalization $\mu(g)$, these expressions are invariant in the
duality transformation $g\to {1\over g}$. This is expected from the duality 
of the S matrix itself - as for the role of the overall normalization, it will
be discussed later.

The conventional normalization \cite{Fring} is $\mu=1$, which corresponds to chosing $<0|\Phi(0)|\a>={1\over \sqrt{2}}$. We shall
make a different choice later on, when we consider the massless limit.

\smallskip

The method can be generalized to models with several types of particles, like the sine-Gordon model: this will be discussed  in the following sections. 
Of course, all the foregoing equations for form-factors have been established within the 
context of ordinary massive integrable field theories. The case
of  massless theories  (first considered in a slightly different context in 
\cite{Aldo})  will
be handled simply by taking the appropriate massless (ultra violet) limit
in all the equations. This is somewhat safer than trying directly 
to formulate 
 axioms  for a massless theory per
se, massless scattering presenting some physical ambiguities (see \cite{Aldophd} for a detailed discussion of this point).

\smallskip

We shall mostly consider
physical properties related with the $U(1)$ currents $\partial\Phi$; hence, our discussion will be centered on the form-factors of this operator. In the last section however, we give the example of a correlator involving vertex operators.

We now discuss in more details the questions at stake in the case of the sinh-Gordon model. 
It has little physical interest in the present 
context, but is pedagogically quite useful

\section{Example: The sinh-Gordon model.}

\subsection{Massless form-factors and the bulk current-current correlators.}

In most of the following calculations, we shall work  in Euclidian space with $x,y$
coordinates. Imaginary
time is at first considered as running along $x$.
The  action and $S$ matrix for the massive sinh-Gordon model were given in the introduction (\ref{shgoract}).

Let us now try to describe the free boson theory as a massless limit of this model.  
First, recall the current correlators (we could use here the notation $\phi,\bar{\phi}$ for the chiral components, but since these get mixed in the presence of  the boundary, we will not do so)
\begeqar
<\partial_z\Phi(z,\bar{z})
\partial_{z'}\Phi(z',\bar{z}')>=
-{2g\over (z-z')^2}\nonumber\\
<\partial_{\bar{z}}\Phi(z,\bar{z})\partial_{\bar{z}'}
\Phi(z',\bar{z}') >=
-{2g\over (\bar{z}-\bar{z}')^2}.
\label{corrshg}
\endeqar
To start, we wish  to recover these correlators using form-factors. We thus take 
the massless limit  $\Lambda\to 0$ of the factorized scattering 
description of (\ref{shgoract}). As discussed in \cite{Houches98}, we start by 
writing
$\alpha=\pm(A+\theta)$ and take simultaneously
$A\to\infty$ and $M\to 0$ with $Me^{A}/2\to m$, $m$
finite. In that case, the spectrum separates into Right  and Left
movers
with respectively $E=P=m e^\theta$ and $E=-P=m e^\theta$.
The scattering of R and L movers is still given by (\ref{smat}) where
$\alpha\to\pm \theta$. The RL and LR scattering becomes a simple phase,
$e^{\mp i\pi B/2}$. This phase will turn out to cancel out at the end
of
all computations, but is confusing to
keep along. We just set it equal to unity in the following, that is
we consider all L and R quantities as commuting.

In this new description of the massless theory, we will need form
factors in order to compute (\ref{corrshg}). 
By taking the massless limit of the formulas \cite{Fring} given in the introduction , it is easy to
check that $\Phi$ can alter only the right or left content of states;
in other words, matrix elements of $\Phi$
between states which have different content both in the
left and right sectors vanish (that is, $\Phi=\phi+\bar{\phi}$!)

Our conventions are conveniently summarized by giving
the one particle form factor of the sinh-Gordon field~:
\begeqar
<0|\Phi(x,y)|\theta>_R&= \mu \exp\left[ me^\t
(x+iy)\right]\nonumber\\
<0|\Phi(x,y)|\t>_L&=\mu \exp\left[me^\t (x-iy)\right],
\label{conv}
\endeqar
and we will use the obvious notation~:
\begeq
f(\t_1,\ldots,\t_{2n+1})=
<0|\Phi|\t_1,\ldots,\t_{2n+1}>_{R,\ldots,R},
\label{nnot}
\endeq
with the normalization of asymptotic states
$_R<\t|\t'>_R=2\pi\delta(\t
-\t')$. Here, $f$ depends on the interaction strength $g$,
but we do not indicate it explicitely for simplicity.
We have the following properties~:
\begeqar
<0|\Phi|\t_1,\ldots,\t_{2n+1}>_{R,\ldots,R}&=& 
\left(<0|\Phi|\t_1,\ldots,\t_{2n+1}>_{L,\ldots,L}\right)^* \nonumber\\
<0|\Phi|\t_1,\ldots,\t_{2n+1}>_{R,\ldots,R}&=&
<0|\Phi|\t_{2n+1},\ldots,\t_1>_{L,\ldots,L}.
\label{propp}
\endeqar
These form factors are expressed just as in the massive case  
\begeq
f(\t_1,\ldots,\t_{2n+1})=\mu
\left({4\sin {\pi B\over 2} \over F_{min}(i\pi,B)}\right)^n
\sigma_{2n+1}^{(2n+1)}P_{2n+1}(x_1,\ldots,x_{2n+1})
\prod_{i<j}{f_{min}(\t_i-\t_j) \over x_i+x_j},
\label{ffexpri}
\endeq
where we introduced $x\equiv e^\t$ and the $\sigma$'s are the  
same symmetric polynomials as before.

We will now chose the normalization $\mu$ by demanding  
that the result (\ref{corrshg}) be recovered. Using form factors, this two point function expands,
assuming $Re\ z<Re\ z'$, as~:
\begeqar
<0|\partial_z\Phi(z,\bar{z})\partial_{z'}
\Phi(z',\bar{z}')|0>=
&-&\sum_{n=0}^\infty \int {d\t_1\ldots d\t_{2n+1}\over
(2\pi)^{2n+1}(2n+1)!}
m^2 \left(e^{\t_1}+\ldots+e^{\t_{2n+1}}\right)^2\nonumber\\
&\times&\exp
\left[m(z-z')\left(e^{\t_1}+\ldots+e^{\t_{2n+1}})\right)\right]
|f(\t_1,\ldots,\t_{2n+1})|^2.
\label{ffnorm}
\endeqar
Now, by relativistic invariance, all the form factors depend
only on differences of rapidities. Setting $m(z-z')\equiv
e^{\t_0}$,
(where $\t_0$ will in general be complex), one
can shift all the  $\beta$'s  by $\t_0$ to factor out, for
any $2n+1$ particle contributions, a factor ${1\over (z-z')^2}$.
Hence,
the form factor expansion gives the result (\ref{corrshg}) provided $\mu$ is
chosen
such that
\begeq
\sum_{n=0}^\infty I_{2n+1}=2g,
\label{niden}
\endeq
where
\begeq
I_{2n+1}= \int {d\t_1\ldots d\t_{2n+1}\over
(2\pi)^{2n+1}(2n+1)!}
 \left(e^{\t_1}+\ldots+e^{\t_{2n+1}}\right)^2
e^{-(e^{\t_1}+\ldots+e^{\t_{2n+1}})}
|f(\t_1,\ldots,\t_{2n+1})|^2.
\label{deff}
\endeq
In practice, this sum cannot be computed analytically, but it
can be easily evaluated numerically. The
convergence is extremely fast with $n$, and for most practical
purposes, the
consideration of up to five particles is enough to get
correct results up to $10^{-4}$.  Similar convergence properties
were observed in \cite{Cardy} in the massive case; note however that in this massless
case, contributions with a higher number of particles are not damped-off 
by exponential mass terms.

It must be emphasized that this result
is very peculiar to the current operator. For most other chiral
operators, the correct $(z-z')$ dependence involves a non trivial
anomalous dimension, instead of the naive engineering dimension.
Hence,
this dependence  is not obtained term by term, as observed here,
but rather once the whole series is summed up. Truncating the series
to
any finite $n$ does not, in such cases, give reliable results all the
way
from short to large distances, unless some additional tricks are performed (see below).

\subsection{Current current correlators with a boundary}

Having fixed the form-factors normalization, let us now consider the
theory with a boundary.  The geometry of the
problem is such that the boundary stands at
$x=0$, and runs
parallel to the $y=it$ axis. The action is now~:
\begeq
S={1\over 16\pi g}\int_{-\infty}^0
dx\int_{-\infty}^\infty dy
\left[
\left(\partial_x\Phi\right)^2
+ \left(\partial_y\Phi\right)^2+\Lambda\cosh \Phi\right]+\lambda
\int_{-\infty}^\infty dy \cosh{1\over 2}\Phi(x=0,y).
\label{bdract}
\endeq
This model is also integrable for any choice of $\Lambda,\lambda$.
The boundary dimension of the
perturbing operator is $x=-g$.
We can in particular take the limit where $\Lambda\to 0$ while
$\lambda$
remains finite.
It then describes a theory which is conformal invariant in the bulk
but has a boundary interaction that breaks this invariance and
induces a flow
from Neumann boundary conditions at small $\lambda$ to Dirichlet
boundary
conditions at large $\lambda$. As discussed in \cite{Houches98} ,  the boundary
interaction is
characterized by an energy scale,
which one can represent as $T_B=me^{\theta_B}$. $T_B$ is related
with the bare coupling in the action (\ref{bdract}) by $\lambda\propto
T_B^{1+g}$.
In the following, since obviously changes of $m$ (which is not a
physical
scale)
can be absorbed in rapidity shifts, we set $m=1$. The effect of the
boundary is
then expressed by the  reflection matrix~:
\begeq
R(\theta)=\tanh \left[{\theta\over 2}-i{\pi\over 4}\right].
\label{rmat}
\endeq

In the picture where imaginary time
is along $x$, the effect of the boundary is represented
by a boundary state. Following \cite{GZ} we can represent
it in terms of the boundary scattering matrix~:
\begeq
|B>=\exp\left[\int_{-\infty}^\infty {d\theta\over 2\pi}
K(\theta_B
-\theta)Z^\dagger_L(\theta)
Z^\dagger_R(\theta)\right]|0>.
\label{Bform}
\endeq
In this formula, $Z^\dagger$ are again the  Zamolodchikov Fateev creation
operators, $K$ is related with the reflection matrix by~:
\begeq
K(\theta)=R\left({i\pi\over 2}-\theta\right)=-\tanh
{\theta\over 2}.
\label{krrel}
\endeq
We do not prove the expression (\ref{Bform}) directly here. It follows however from
the compatibility between the present computation, and the one we will do next,
in a modular transformed point of view where the direction of imaginary time will have been switched.

One can expand the boundary state into the convenient form~:
\begeqar
|B>&=&\sum_{n=0}^\infty \int_{0<\t_1<\ldots<\t_n} {d\theta_1\over 2\pi}\ldots
{d\theta_n\over2\pi}
K(\t_B-\t_1)\ldots K(\t_B-\t_n)\nonumber\\
&\times&Z^\dagger_L(\t_1)\ldots
Z^\dagger_
L(\t_n)
Z^\dagger_R(\t_1)\ldots Z_R^\dagger(\t_n)|0>.
\label{Bformconv}
\endeqar
Here, we have ordered the rapidities, but the result is the same 
as the unordered integral with the additional symmetry factor $1/n!$, 
the contributions from L and R scattering cancelling out after reordering. 
Observe now that by analyticity, the matrix elements of
$\partial_z\Phi$
between the ground state and any state with at least one L moving
particle are identically zero.  More generally,  the only
non vanishing matrix elements of $\partial_z\Phi$ are those where bra
and ket
have the same L moving part. The same results apply by exchanging
$\partial_z$
with $\partial_{\bar{z}}$ and L with R moving particles. As a result
one gets immediately two of the four current correlators~:
\begeqar
<\partial_z\Phi(z,\bar{z})
\partial_{z'}\Phi(z',\bar{z}')>&=
-{2g\over (z-z')^2}\nonumber\\
<\partial_{\bar{z}}\Phi(z,\bar{z})\partial_{\bar{z}'}
\phi(z',\bar{z}')>&=-{2g\over (\bar{z}-\bar{z}')^2},
\label{trivial}
\endeqar
which are identical with the ones without a boundary.

The two other correlators are more difficult to
get. Let us consider for instance~:
\begeq
<0|\partial_{\bar{z}}\Phi(z,\bar{z})
\partial_{z'}\Phi(z',\bar{z}')|B>.
\label{tobf}
\endeq
The first non trivial contribution comes
from the two particle term in the expansion of the boundary state~:
\begeqar
\int_{-\infty}^\infty &&{d\t\over 2\pi}K(\t_B-\t) <0|
\partial_{\bar{z}}\Phi(z,\bar{z})
\partial_{z'}\Phi(z',\bar{z}')Z^\dagger_L(\t)Z^\dagger_R(\t)|0>=\nonumber\\
&\times&\mu^2\int_{-\infty}^\infty {d\t\over 2\pi} K(\t_B-\t)
e^{2\t}
\exp\left[ e^\t (\bar{z}+z')\right].
\label{first}
\endeqar
More generally, because $|B>$ is a superposition of states with
equal numbers of left and right moving particles, and
$\partial_z\phi$,
respectively $\partial_{\bar{z}}\phi$ act only on R, respectively L,
particles,
the expansion of (\ref{tobf}) takes a very simple form~:
\begeqar
\sum_{n=0}^\infty \ \int &&{d\t_1\ldots d\t_{2n+1}\over
(2\pi)^{2n+1}(2n+1)!}
K(\t_B-\t_1)\ldots K(\t_B-\t_{2n+1})
\left(e^{\t_1}+\ldots+
e^{\t_{2n+1}}\right)^2
\nonumber\\
&&\exp
\left[(\bar{z}+z')\left(e^{\t_1}+\ldots+e^{\t_{2n+1}}\right)
\right]
|f(\t_1,\ldots,\t_{2n+1})|^2.
\label{step}
\endeqar
This correlation function depends on the product $e^{\theta_B}
(\bar{z}+z')$. It
is scale invariant at the UV and IR fixed point. These correspond
respectively
to sending $\theta_B$ to $\mp \infty$, that is  the coupling
$\lambda$ in the action
to $0$ or $\infty$, in other words Neumann  or Dirichlet boundary
conditions.
In the first case, $K=1$, in the second, $K=-1$. Comparing with
(\ref{ffnorm}) and
(\ref{niden}) we find, as expected, that~:
\begeq
<0|\partial_{\bar{z}}\Phi(z,\bar{z})
\partial_{z'}\Phi(z',\bar{z}')|B>=\pm {2g\over(\bar{z}+z')^2},
\label{res}
\endeq
for Neumann, respectively Dirichlet boundary conditions.
Although trivial, this result shows that the form factor
expansion is well behaved, and allows us to study the correlator
all the way from the UV to the IR fixed point when there is a
boundary perturbation.  In figures 1 and 2 we show the
one particle (which is independent of $B$)
and three particles contributions.  We observe
that indeed the convergence, by looking at the respective
contributions, is very rapid.

\begin{figure}[tbh]
\centerline{\psfig{figure=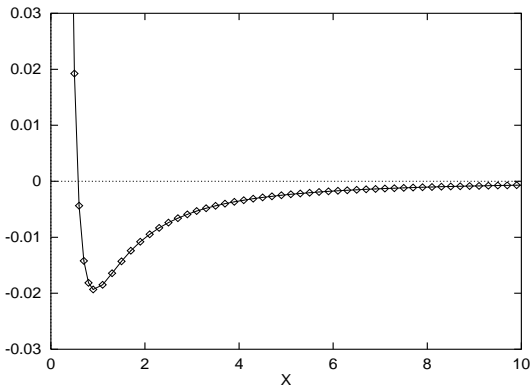,height=2.in,width=3.in}}
\caption{One particle contribution.}
\end{figure}

\begin{figure}[tbh]
\centerline{\psfig{figure=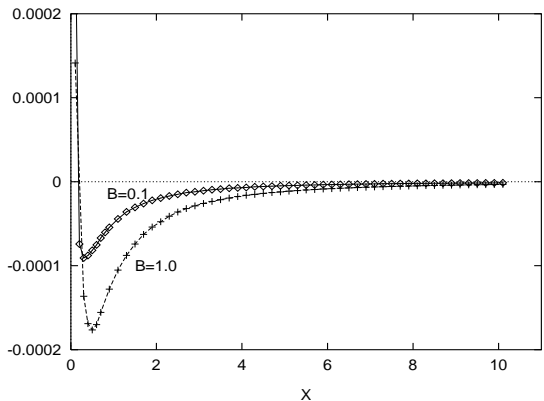,height=2.in,width=3.in}}
\caption{Three  particles contribution for $B=1,0.1$.}
\end{figure}

The only drawback of this expansion is that it  is not suited
for studying the correlation of two operators right at the boundary.
Indeed
in that case, $Re (\bar{z}+z')=0$, and the integrals in (\ref{step}) do not
converge.
To solve this problem, we can introduce a modular transformed
picture. We now
consider
the imaginary time as running along the $y$ axis. Now the boundary
is not represented as a state; rather, the whole space of states is
different,
since now we have only a half space to deal with. The asymptotic
states are
not pure L or R moving, but are mixtures. For instance,  one particle
states are~:
\begeq
||\t>=|\t>_R+R(\t)|\t>_L.
\label{asym}
\endeq
More generally, asymptotic states are obtained by adding to
$|\t_1,\ldots,\t_n>_{R,\ldots,R}$ all combinations with
different
choices of $R$ particles transformed into $L$ particles, via action
of the
boundary. Only the following two terms contribute~:
\begeq
||\t_1,\ldots,\t_n>=
|\t_1,\ldots,\t_n>_{R\ldots ,R}
+\ldots+R(\t_1)\ldots
R(\t_n)|\t_n,\ldots,\t_1>_{L,\ldots,L}+\ldots.
\label{asymr}
\endeq
Although we used the same notation as previously,  different things
are meant by L,R. To make it clear, we now use the conventions~:
\begeqar
<0|\Phi(x,y)|\theta>_R
=\mu\exp[me^\theta(-y+ix)]\nonumber\\
<0|\Phi(x,y)|\theta>_L=\mu\exp[me^\theta(-y-ix)].
\label{newconv}
\endeqar
To keep the notations as uniform as possible, we introduce the new
coordinates~:
\begeq
w(z)\equiv iz=-y+ix,
\label{defdef}
\endeq
so here R movers depend on $w$, L movers on $\bar{w}$. The
normalization $\mu$
is of course the same as before, and as before the LL and RR
correlators
do not depend on the boundary interaction. One finds~:
\begeqar
<0|\partial_w\Phi(w,\bar{w})
\partial_{w'}\Phi(w',\bar{w}')|0>&=
-{2g\over (w-w')^2}\nonumber\\
<0|\partial_{\bar{w}}\Phi(w,\bar{w})\partial_{\bar{w}'}
\Phi(w',\bar{w}')|0>&=-{2g\over (\bar{w}-\bar{w}')^2},
\label{trivialbis}
\endeqar
where we used the fact that $|R(\theta)|^2=1$. When compared with
(\ref{trivial}),
these
 correlators have an overall minus sign
due to the dimension $h=1,\bar{h}=0$ (resp. $h=0,\bar{h}=1$)
of the operators.

Let us now consider~:
\begeq
<0|\partial_{\bar{w}}\Phi(w,\bar{w})\partial_{w'}
\Phi(w',\bar{w}')|0>.
\label{tbdbis}
\endeq
To compute it, we insert a complete set of states which are of the
form (\ref{asym}). In the
massless case however, since $\partial_w\Phi$ is a R operator,
 $\partial_{\bar{w}'}\Phi$ a L  operator, the only terms that
contribute are in fact the ones with either all L or all R moving
particles, as written in (\ref{asymr}). Thus, (\ref{tbdbis}) expands simply as~:
\begeqar
-\sum_{n=0}^\infty \int && {d\theta_1\ldots
d\theta_{2n+1}\over
(2\pi)^{2n+1}(2n+1)!}R(\theta_1-\theta_B)\ldots
R(\theta_{2n+1}-\theta_B)\left(e^{\theta_1}+
\ldots+e^{\theta_{2n+1}}\right)^2\nonumber\\
&&\exp\left[(\bar{w}-w')(e^{\theta_1}+\ldots+e^{\theta_{2n+1}})\right]
|f(\theta_1,\ldots,\theta_{2n+1})|^2.
\label{newexp}
\endeqar
Observe the crucial   minus sign when compared to (\ref{step}). It occurs
because in one geometry the correlator depends on $\bar{z}+z'$, while
in the other on $\bar{w}-w'$.  This now converges provided  $y>y'$,
even if $x=x'=0$
ie the operators are sitting right on the boundary. Now, using the
fact
that from factors depend only on differences of rapidities,  this
expression
can be mapped with (\ref{step}) if we formally set
$\theta=\theta'+i{\pi\over 2}$, provided one has~:
\begeq
K(\theta)=R\left(i{\pi\over 2}-\theta\right),
\label{idenden}
\endeq
as claimed above.

To summarize, we can write the left right current current correlator
in two
possible ways. By using the boundary state one finds~:
\begeq
<\partial_{\bar{z}}\Phi(x,y)\partial_{z'}\Phi(x',y')>=
\int_0^\infty dE \ {\cal G}(E) \exp\left[E(x+x')-iE(y-y')\right],
\label{idenmaktata}
\endeq
(recall that $x,x'<0$). One obtains ${\cal G}(E)$ simply by fixing
the energy to
a particular value in (\ref{step}). When this is done, the remaining
integrations
occur on a finite domain for each of the individual particle energies
since $\sum_{i=1}^{2n+1} e^{\t_i}=E$, and there is no problem of
convergence
anymore. One then gets~:
\begeqar
{\cal G}(E)
&= &\sum_{n=0}^\infty \int_{-\infty}^{\ln E}
{d\t_1\ldots
d\t_{2n}\over(2\pi)^{2n+1}(2n+1)!}{E^2\over E-e^{\t_1}-\ldots
-e^{\t_{2n}}}\nonumber\\ &\times&
K(\t_B-\t_1)\ldots K(\t_B-\t_{2n})
K\left[\t_B-\ln\left(E-e^{\t_1}-
\ldots-e^{\t_{2n}}\right)\right]
\nonumber\\ &\times&
\left|f\left[\t_1\ldots\t_{2n},\ln\left(E-e^{\t_1}-\ldots
-e^{\t_{2n}}
\right)\right]\right|^2,
\label{nicenice}
\endeqar
with the constraint $\sum_{i=1}^{2n} e^{\t_i}\leq E$.
The denominator might suggest some possible divergences; it is
important
however to realize that it vanishes if and only if
the particle with rapidty $\t_{2n+1}$ has vanishing energy,
in which case the form factor vanishes too. We can now shift
the integrands to write equivalently~:
\begeqar
{\cal  G}(E)&=&E
 \sum_{n=0}^\infty \int_{-\infty}^0 {d\theta_1\ldots d\theta_{2n}
\over(2\pi)^{2n+1}(2n+1)!}{1\over 1-e^{\theta_1}-\ldots
-e^{\theta_{2n}}}\nonumber\\ &\times&
K(\ln (T_B/E)-\theta_1)\ldots K(\ln (T_B/E)-\theta_{2n})
K\left[\ln
(T_B/E)-\ln\left(1-e^{\theta_1}-\ldots-e^{\theta_{2n}}\right)\right]\nonumber\\
&\times& \left|f\left[\theta_1\ldots\theta_{2n},
\ln\left(1-e^{\theta_1}-
\ldots-e^{\theta_{2n}}\right)\right]\right|^2,
\label{nicenicenicetata}
\endeqar
where the constraint
$\sum_{i=1}^{2n} e^{\t_i}\leq 1$ is implied,  we used the fact
that form-factors depend only on rapidity differences,
and
$T_B\equiv e^{\t_B}$.

By using the dual picture, one finds, instead of (\ref{idenmaktata})~:
\begeq
<\partial_{\bar{z}}
\Phi(x,y)\partial_{z'}\Phi(x',y')>=
\int_0^\infty dE {\cal F}(E) \exp\left[-iE(x+x')-E(y-y')\right],
\label{idenmaktiti}
\endeq
where~:
\begeqar
{\cal  F}(E)&=&-E
 \sum_{n=0}^\infty \int_{-\infty}^0 {d\theta_1\ldots d\theta_{2n}
\over(2\pi)^{2n+1}(2n+1)!}{1\over 1-e^{\theta_1}-\ldots
-e^{\theta_{2n}}}\nonumber\\ &\times&
R(\theta_1-\ln (T_B/E))\ldots R(\theta_{2n}-\ln (T_B/E))
R\left[\ln\left(1-e^{\theta_1}-\ldots-e^{\theta_{2n}}\right)-\ln
(T_B/E)
\right]\nonumber\\ &\times&
\left|f\left[\theta_1\ldots\theta_{2n},
\ln\left(1-e^{\theta_1}-
\ldots-e^{\theta_{2n}}\right)\right]\right|^2,
\label{nicenicenicetiti}
\endeqar
where in (\ref{nicenicenicetata}) and (\ref{nicenicenicetiti}) the constraint
$\sum_{i=1}^{2n} e^{\t_i}\leq 1$ is implied.
The two expressions are in correspondence by the simple analytic
continuation~:
\begeq
{\cal G}(E)= i{\cal F}(iE).
\label{matsu}
\endeq

\section{The sine-Gordon Model.}

In this section we follow the same line of thought for the
sine-Gordon model.
This is the massive deformation of the free boson  which preserves
integrability with either  boundary
interactions  used in the fractional quantum Hall problem {\it and} the
anisotropic
Kondo model \cite{Houches98}.  Thus the form factors of the sine-Gordon model in the
massless limit will be the quantities we need.
The solitons/anti-solitons and breathers quasi-excitations make the
problem
more complicated but the results presented before hold with the
addition
of a few indices (and rather more complicated form factors).

We start as before with the  massive sine-Gordon model, whose 
action reads~:
\begeq
S={1\over16\pi g}\int_{-\infty}^\infty dxdy \ \left[
\left(\partial_x\Phi\right)^2
+ \left(\partial_y\Phi\right)^2+\Lambda\cos \Phi\right].
\label{shgoracti}
\endeq
Notice again that we have used a different normalization than in \cite{Houches98}: this will
avoid carrying factors of $2\pi$ all along the following paragraphs. Recall that  $g={\beta^2\over 8\pi}$, $\beta$ 
the usual sine-Gordon parameter.

The form
factors approach is formally the same, albeit more complicated
because the particle content is much richer, and depends on $g$.
For $1/2 < g <1$, only solitons/anti-solitons appear in the spectrum
of the theory.  This is the so called repulsive case, with $g=1/2$
the free fermion point (giving rise, in quantum impurity problems,
to so called ``Toulouse limits''.  When $0<g<1/2$, the particle content is enriched
by $[1/g-2]$ bound states, called breathers.  In the following
we will denote by the indices $a=\pm$ the solitons and
anti-solitons, and $a=1,2,...,[1/g-2]$ the breathers.
The solitons form factors in the massive case were written by
Smirnov \cite{Smirnov}
and we obtain the massless form factors by taking the appropriate
limit of the massive ones.  Only right and left moving form
factors survive in this limit, as in the sinh-Gordon case.
Moreover, the symmetry of the action dictates that only form factors
with
total topological charge zero are non-zero for the current
operator. As an example, the soliton/anti-soliton form factor
is given by~:
\begeqar
<0|{1\over 2\pi}\partial_z\Phi(z,\bar{z})|\t_1,\t_2>_
{aa'}^{RR}&=&
a'\mu m  d\ 
e^{(\t_1+\t_2)/2}\nonumber\\
&\times&{\zeta(\t_1-\t_2)\over
\cosh {(1-g)\over 2g}(\t_1-\t_2+i\pi)}\exp\left[m(e^{\t_1}
+e^{\t_2})z\right],
\label{ffsmir}
\endeqar
with $a+a'=0$
and $a=\pm$ stands for soliton (resp. antisoliton).
From \cite{Smirnov} one has~:
\begeq
\zeta(\t)=c \sinh{\t\over 2} \exp\left(\int_0^\infty
{\sin^2 {x(i\pi-\t)\over 2\pi} \sinh { (1-2g)x\over 2(1-g)}\over x
\sinh { g x\over 2(1-g)}\sinh x\cosh { x \over 2}}
dx\right),
\label{zzzzz}
\endeq
(this is essentially the minimum form-factor discussed in the introduction) with the constant $c$ given by~:
\begeq
c=\left( {4(1-g)\over g}\right)^{1/4}\exp\left({1\over
4}\int_0^\infty
{\sinh {x \over 2} \sinh { (1-2g)x\over 2(1-g)}\over x
\sinh { g x\over 2(1-g)}\cosh^2 { x \over 2}} dx\right),
\label{constc}
\endeq
and $d$ by~:
\begeq
d={1\over 2\pi c} {(1-g)\over g},
\label{ddd}
\endeq
The  normalization constant $\mu$ can be determined from first principles. 
 Indeed,
the operator $\partial_x\Phi$ being related with the $U(1)$ charge, we
need
that
\begeq
^+_R<\theta_1|\int_{-\infty}^\infty
{1\over 2\pi}\partial_x\Phi|\theta_2>_+^R\ = 2\pi
\delta(\theta_1-\theta_2),
\label{charr}
\endeq
using the fact that a soliton for the bulk theory (\ref{shgoracti}) obeys
$\Phi(\infty)-\Phi(-\infty)=2\pi$. On the other hand, using the dependence 
of the form-factor on spatial coordinates, 
the left hand side is 
\begeq
^+_R<\theta_1|{1\over 2\pi}\partial_x\Phi|\theta_2>_+^R\ \int dx e^{im(e^{\theta_1}-e^{\theta_2})x}={2\pi\over me^{\theta}}\ 
^+_R<\theta_1|{1\over 2\pi}\partial_x\Phi|\theta_1>_+^R \delta(\theta_1-\theta_2),
\label{charrr}
\endeq
Comparing the two and using crossing leads to the identity 
\begeq
i\mu  d\zeta(i\pi)=1
\endeq
or 
\begeq
\mu=2\pi{g\over 1-g}={1\over cd}\label{valofmu}.
\endeq

The other soliton-antisoliton form  factors follow from the sophisticated  analysis
of \cite{Smirnov}. 
Their expression simplifies in the case $g={1\over t}$, $t$ an
integer. This is the physically
relevant case for the $\nu={1\over t}$ fractional quantum Hall
effect.
One then finds~:
\begeqar
<0|{1\over 2\pi}\partial_z\Phi(z,\bar{z})|\t_1,\ldots,
\t_{2n}>^{-\ldots -,+\ldots +}_{R\ldots R}=\mu m (2 d)^n
e^{(\t_1+\ldots+\t_{2n})/2}\prod_{i<j}\zeta(\t_i-\t_
j)\nonumber\\
\sinh \left[{t-1\over 2} \sum_{p=1}^n
(\t_{p+n}-\t_p-i\pi)\right]\prod_{p=1}^n\prod_{q=n+1}^{2n}
\sinh^{-1} (t-1)
(\t_q-\t_p) \ det H.
\label{ffsmirbis}
\endeqar
The matrix $H$ is obtained as follows. First introduce  the
function~:
\begeq
\psi(\alpha)=2^{t-2}\prod_{j=1}^{t-2}\sinh
{1\over 2}\left(\alpha-i{\pi j\over t-1}+i{\pi\over 4}\right).
\label{interm}
\endeq
One then defines the matrix elements as~:
\begeq
H_{ij}={1\over 2i\pi}\int_{-2i\pi}^0 d\alpha
\prod_{k=1}^{k=2n}
\psi(\alpha-\beta_k)\exp\left[(n-2j-1)
\alpha+(n-2i)(t-1)\alpha\right],
\label{matrix}
\endeq
where $i,j$ run over $1,\ldots, n-1$.  It is not difficult to
convince
oneself that this produce a symmetric polynomial of the right degree.
Although cumbersome, it is an easy task to extract these
determinants, as examples we find for $g=1/3$~:
\begeqar
det H=\exp\left(-{1\over 2} \sum_{i=1}^{2n} \t_i\right) \
\sigma_1(e^{\t_p}) , \ \ n=2, \nonumber\\
det H=\exp(-\sum_{i=1}^{2n} \t_i) \ \sigma_1(e^{\t_p})
\sigma_3(e^{\t_p}), \ \ n=3,
\label{exdet}
\endeqar
up to irrelevant phases and
with the $\sigma_q$'s defined previously.  Having these expression
we can get all form factors using the axiomatics sketched in the introduction.
 For example, the solitons form factors with different
positions
of the indices $a_i$ we use the symmetry property (\ref{genffeq}):
\begeqar
&f(\t_1,...,\t_i,\t_{i+1},...,\t_n)_{a_1,...,
a_i,a_{i+1},...,a_n}=&
\nonumber\\ 
&f(\t_1,...,\t_{i+1},\t_i,...,\t_n)_{a_1,...,
b_{i+1},b_i,...,a_n} S^{b_i,b_{i+1}}_{a_i,a_{i+1}}
(\t_i-\t_{i+1}) .&
\label{commeps}
\endeqar
Here again, we omit the distinction between left and right moving
form
factor, they are simply related by complex conjugation.
At the points $g=1/t$ the soliton $S$ matrix used in the last
expression is reflectionless and basically just permutes the
rapidities up to a phase.  When there are breathers, the
soliton $S$ matrix has poles corresponding to the bound
states at the points $\t=iu_{ab}^c= i\pi-{i\pi g\over (1-g)} m$ for the
$m$'th breather.
In view of the last relation, this induces poles
in the form factors.  We obtain the breather form factors from
these poles~:
\begeqar
i{\rm res}_{\t_{n-1}-\t_n=iu_{ab}^c} f(\t_1,...,\t_{n-1},\t_n)_{a_1,...,
a_{n-1},a_n}&=&r_m (-1)^{{a_n+1\over 2}m}
\delta_{a_{n-1}+a_n} \nonumber\\
&\times& f(\t_1,...,\t_{n-1}+{i\pi\over 2}-i{i\pi gm\over 2 (1-g)})_{
a_1,...,a_{n-2},m},
\label{polerel}
\endeqar
and $r_m$ is given by the residue at $iu_{ab}^c=i\pi-{i\pi g\over (1-g)}
m$~:
\begeq
r_m=\left[S_{++}^{++}\left(i\pi -{i\pi mg\over 1-g}\right) { g\over (1-g)}
\sin\pi {1-g\over g}\right]^{1/2}.
\label{poledeux}
\endeq
Having these relations, we posess all ingredients to compute all
form factors for $g=1/t$.  Using them for the computation
of the current correlations is then merely an extension of the previous
results for sinh-Gordon with indices.  The normalisation of the
form factors, $\mu$, should also ensure that  (\ref{trivial}) is reproduced.
This is fixed by
introducing a complete basis of states~:
\begeq
1=\sum_{n=0}^\infty \sum_{a_i} \int {d\t_1 ...
d\t_n\over
(2\pi)^n n!} |\t_1,...,\t_n>_{a_1,...,a_n}
{}^{a_n,...,a_1}<\t_n,...,\t_1|
\label{compbas}
\endeq
and computing the correlations exactly like in the sinh-Gordon case.

Keeping a finite number of form-factors and demanding that the two point function
is properly normalized gives rise to values of $\mu$ which are slightly different from
(\ref{valofmu}). How different is a good measure of the convergence of the expansion, and 
the validity of the truncation. For $g=1/3$, the one breather and
2 solitons form factors normalise to $\mu=3.14$ which is very
close to the exact $\pi$.  Similarly for $g=1/4$ we found
from the contributions up to two solitons that $\mu=2.05$
to compare with $2.094=2\pi/3$.

Calculations in the presence of an integrable boundary  interaction are also
done like in the sinh-Gordon case. The boundary state is now 
given by~:
\begeqar
|B>=\sum_{n=0}^\infty && \int_{0<\t1<...<\t_n}
{d\theta_1\over 2\pi}\ldots {d\theta_n\over 2\pi}K^{a_1b_1}(\t_B-\t_1)...K^{a_nb_n}(\t_B-\t_n)\nonumber\\
&\times &Z_L^{\dagger a_1}(\t_1)\ldots Z_L^{\dagger a_n}(\t_n) Z_R^{\dagger b_1}(\t_1)...Z_R^{\dagger b_n}(\t_n),
\label{bstate}
\endeqar
with an implicit sum on the indices.
The matrix $K^{ab}$ is related to the boundary $R$ matrix
in the following way~:
\begeq
K^{ab}(\t)=R^a_{\bar{b}} \left(i{\pi\over 2}-\t\right).
\label{relkr}
\endeq
The $\bar{b}$ means that we take the conjugate of the indices
ie. $\pm\rightarrow \mp$ and $m\rightarrow  m$.

From the previous expressions, we can compute de current-current
correlation function in the presence of a boundary for $g=1/t$.
The results we will get depends on the boundary interaction, in the
next subsection we
 present some results    for the boundary sine-Gordon model, which  is of
relevance to tunneling experiments  in fractional  quantum Hall devices
 \cite{Houches98,KaneFisher,Moon,FLS}.

\section{Conductance in the fractional quantum Hall effect}

\subsection{General remarks.}

The boundary sine-Gordon action is 
\begeq
S={1\over 16\pi g}\int_{-\infty}^0
dx\int_{-\infty}^\infty dy
\left[
\left(\partial_x\Phi\right)^2
+ \left(\partial_y\Phi\right)^2\right]+\lambda
\int_{-\infty}^\infty dy \cos{1\over 2}\Phi(x=0,y).
\label{bdracti}
\endeq
The reflection matrices have been worked out in \cite{GZ,FSW}. For generic values of the coupling $g$,
the amplitude for the processes $+\to +$ and $-\to -$ is
$R^\pm_\pm(\t-
\t_B)$, and for the processes $+\to -$ and $-\to +$ it is $
R^\pm_\mp(\t-\t_B)$
with (the notation is slightly changed with respect to \cite{Houches98}, where 
outgoing indices were not raised):
\begeqar
R^\pm_\mp(\theta)&={e^{(1-g)\theta\over
2g}}R(\theta)\nonumber\\
R^\pm_\pm(\theta)&= i{e^{(g-1)\theta\over 2g}}R(\theta)
\label{bdrs}
\endeqar
where the function $R$ reads~:
\begeqar
R(\theta)&={e^{i\gamma}
\over 2\cosh\left[{(1-g) \theta\over 2g}-i{\pi\over 4}\right] }
\prod_{l=0}^\infty{Y_l(\theta)\over Y_l(-\theta)}\nonumber\\
Y_l(\theta)&={\Gamma\left({3\over 4}+l{(1-g)\over g}
-{i(1-g)\theta\over 2\pi g}\right)\Gamma\left({1\over
4}+(l+1){(1-g)\over g}
-{ i(1-g)\theta\over 2\pi g}\right)\over \Gamma\left({1\over
4}+(l+1/2)
{(1-g)\over g} -{(1-g) i\theta\over 2\pi g}\right)\Gamma\left({3\over
4}+(l+1/2)
{(1-g)\over g} -{ i(1-g)\theta\over 2\pi g}\right)}.
\endeqar
In (\ref{bdrs}), our conventions are such that
in the UV limit ($\t_B\to-\infty$) the scattering is totally
off-diagonal so a soliton bounces back as an anti-soliton, in
agreement with classical limit results for Neumann boundary
conditions. A useful integral representation of $R$ is given by~:
\begeq
R(\theta)={e^{i\gamma}
\over 2\cosh\left[{(1-g) \theta\over 2g}-i{\pi\over 4}\right] }
\exp\left( i\int_{-\infty}^{\infty} {dy\over 2y} \sin{2(1-g) \theta
y\over g\pi}
{\sinh({1-2g\over g})y \over \sinh 2y\cosh {(1-g) y\over g}
}\right).
\label{norm}
\endeq
Recall that the spectrum is made of one breather and the pair soliton
antisoliton in the whole domain $1/3\leq g <1/2$.
More breathers appear
for $g<1/3$ (and  the reflection matrix of the 1- breather
is always the same as in the sinh-Gordon case.)
There are no breathers for $g>1/2$.

The physical quantity of interest in this case corresponds to 
the AC conductance at vanishing temperature in the edge states tunneling problem:
this problem has been described in details in \cite{Houches98}, to which we refer the reader.

A standard way of representing the conductance  is through the Kubo formula~\cite{Wong}:
\begeq
G(\omega_M)=-{1\over 8\pi \omega_M L^2}\int_{-L}^L
dx \ dx'\int_{-\infty}^\infty dy \ e^{i\omega_M y}<j(x,y)
j(x',0)>,\label{kubozero}
\endeq
where $\omega_M$ is a Matsubara frequency, $y$ is imaginary time,
$y=it$.
One gets back to real physical frequencies by letting
$\omega_M=-i\omega$.
In (\ref{kubozero}) , $j$ is the physical current in the unfolded system  
\cite{Houches98}.
  Without impurity, the AC
conductance of the Luttinger liquid is frequency independent, $G=g$.
When adding the impurity, it becomes $G={g\over 2}+\Delta G$. After
some
simple manipulations using the folding \cite{Houches98}, one finds~:
\begeqar
\Delta G(\omega_M)&=&{1\over 8\pi \omega_M
L^2}\int_{-L}^0
dxdx'\int
_{-\infty}^\infty dy e^{i\omega_M
y} \nonumber\\
 &\times&\left[<\partial_z\Phi(x,y)\partial_{\bar{z}'
}\phi(x',0)>
+<\partial_{\bar{z}}\Phi(x,y)\partial_{z'}\Phi(x',0)>\right],
\label{kubozeri}
\endeqar
where $z=x+iy$. The strategy is simply to evaluate the current-current correlator  
 using form-factors, and extract the conductance from (\ref{kubozeri}).

We will do so for special values of $g$, but first, we  can 
extract some general  features of the UV and IR expansions easily. To do so, consider
the soliton antisolitons reflection matrix. Evaluating the integral
in \ref{norm} by the residues method leads to a double expansion
of the elements $R^\pm_{\pm}$ in powers of $\exp(\t)$
and
$\exp({1\over g}-1)\theta$ .
This leads for the conductance to a  double power series
in $(\omega/T_B)^{-2+2/g}$  and $(\omega/T_B)^2$ in the IR,
$(T_B/\omega)^{2-2g}$  and $(T_B/\omega)^{2}$ in the UV. Breathers
do not change this result.
For instance for the 1-breather, since the reflection matrix is the
 same as in the sinh-Gordon case,
and therefore $g$ independent, the  contributions expand
as a series in $(\omega/T_B)^2$ in the IR,  $(T_B/\omega)^{2}$ in the
UV.
This holds for any coupling $g$. Therefore, as
first argued by  Guinea et al. \cite{Guineaetal},  at low
frequency,
the conductance goes as $\omega^2$ for $g<1/2$, $\omega^{-2+2/g}$ for
$g>1/2$.
The $\omega^2$ power would seem
to indicate that there should be a $T^2$ term in the DC conductance,
but this
is not correct because only the modulus square of
$R^\pm_{\pm}$
contribute to the DC conductance, and this expands only as powers
of $\exp({1\over g}-1)\beta$ .

The presence of analytic terms in the IR is a straightforward
consequence of the fact that IR perturbation theory involves
an infinity of counter-terms, in particular polynomials in
derivatives of $\Phi$ \cite{Guineaetal}. More surprising maybe is the fact
that we find analytical terms in the UV.
This requires some discussion. The UV terms follow from the short
distance behaviour of the correlation function of the current. For
any
operator
$O$ we could write formally,
\begeq
<O(x',y')O(x,y)>_\lambda=\sum_{n=0}^\infty (\lambda)^{2n}
\int d1\ldots dn <O(x',y')O(x,y)\cos{1\over
2}\Phi(1)\ldots
\cos{1\over 2}\Phi(n)>_{\lambda=0}.
\label{peda}
\endeq
From
(\ref{peda}), one would naively expect that the two point function of the
current expands as a power series in $\lambda^2$, which would lead
to a power series in $(\omega/T_B)^{2g-2}$. This is incorrect
however because, even if
 integrals are convergent at short distance for $g<1/2$, they are
always  divergent
at large distances. It is  known that these IR
divergences give precisely
 rise to non analyticity in the coupling constant $\lambda$.
One usually writes~:
\begeq
<O(x',y')O(x,y)>=\sum_i C^i_{OO}(x'-x,y'-y) O_i(x,y),
\label{pedai}
\endeq
where $O_i$ are a complete set of local operators in the theory
and the $C$'s are structure functions. These, being local quantities,
have
 analytic behaviour
in $\lambda$.  However, $<O_i(x,y)>$ being non local is in
 general non analytic - actually, on dimensional grounds,
\begeq
<O_i(x,y)>\propto\lambda^{\Delta/(1-g)}\propto T_B^{\Delta},
\label{pedaii}
\endeq
where $\Delta=h+\bar{h}$ is the (bulk) dimension of the field $O_i$.
If we
   computed the conductance perturbatively using Matsubara formula,
we  would
 use (\ref{peda}) with
$O$ the electrical current operator.  The case $O_i$ the identity
operator
gives rise to an analytical expression in $\lambda$, but eg the
 case $O_i=\partial_z\partial_{\bar{z}}\Phi$ gives
$\lambda^{2/(1-g)}$ times
 an analytical expression in $\lambda$ (its mean value can be non
zero because there is a boundary). More generally, since the only
operators $O_i$ appearing in the case of the current
are polynomials in derivatives of $\Phi$, all with integer
dimensions, we
expect that the two point function of the current will expand as a
 double series of the form $\lambda^{2n}\lambda^{2m/(1-g)}$, ie going
back
 to $T_B$ variable
, that the conductance will expand as a double series of the form
 $(T_B/\omega)^{2n(1-g)}(T_B/\omega)^{2m}$, in agreement with
the form factors result.

\subsection{The free case.}

In the case $g=1/2$ one has simply~:
\begeqar
R^\pm_\mp(\theta)=P(\theta)&={e^\theta\over
e^\theta +i}\nonumber\\
R^\pm_\pm(\theta)=Q(\theta)&={i\over e^\theta+i}.
\label{freeref}
\endeqar
In that case, only the soliton-antisoliton form factor is non zero,
$f(\theta_1,\theta_2)=i\mu e^{\theta_1/2}e^{\theta_2/2}$,
with the normalization $\mu=2\pi$ and we have set $m=1$.
Hence, ${\cal F}(\omega)$ is readily evaluated
\begeq
{\cal F}(\omega)=\int_{-\infty}^\infty
\int_{-\infty}^\infty
d\theta_1d\theta_2\delta(e^{\theta_1}+e^{\theta_2}-\omega)
[Q(\theta_1)Q(\theta_2)-P(\theta_1)P(\theta_2)]e^{\theta_1}e^{\theta_2},
\label{freef}
\endeq
so
\begeq
{\cal F}(\omega)=\omega\int_0^1 dx {x(1-x)+1\over
\left(x+i{T_B\over\omega}\right)
\left(\omega-x+i{T_B\over\omega}\right)},
\label{freefi}
\endeq
from which it follows that
\begeq
\Delta G(\omega)={1\over 4}-{T_B\over
2\omega}\tan^{-1}(\omega/T_B).
\label{freefii}
\endeq
Thus we find
\begeq
G(\omega)={1\over 2}\left(1-{T_B\over \omega}
\tan^{-1}(\omega/T_B)\right).
\label{freefiii}
\endeq
This is in agreement with the solution of \cite{KaneFisher}.

\subsection{$G(\omega)$ at $g=1/3$.}

The conductance for $g=1/3$ has a direct application to the
quantum Hall effect.  Comparing with the free case, previously
treated, we now have a breather in the spectrum and non zero
form factors for all number of rapidities.  Still the convergence
is such that evaluating the first few form factors give results
to a very good accuracy, independently of the regime, UV or IR,
in which we make the computation.

In this case, the first few non zero form factors are
$f_1$, $f_{\pm,\mp}$, $f_{\pm,\mp,1}$, $f_{1,1,1}$, etc...
Here the subscript ``1" denotes the breather.  The first step is
to compute the normalisation in order to satisfy (\ref{corrshg}).
When computing this normalisation,
we find that the first two form factors account
for the whole result to more than one percent accuracy.  Then
including
the 1 breather-2 solitons form factor is sufficient to get
the result to a very good accuracy ($f_{1,1,1}$ is
negligeable). Actually one observes that the speed of  convergence
of the form factor expansion varies  geometrically  with the number
of solitons
(counting the breathers as 2 solitons).

In order to get the
conductance we need the reflection matrices, they were given
in previous expressions and reduce to a simpler form for this value
of $g$~:
\begeq
R(\theta)={1\over 2 \cosh(\theta-{i\pi\over 4})}
{\Gamma(3/8-{i\theta\over 2\pi})\Gamma(5/8+{i\theta\over 2\pi})
\over \Gamma(5/8-{i\theta\over 2\pi})\Gamma(3/8+{i\theta\over 2\pi})}
\label{rgtiers}
\endeq
and the breather reflection matrix is~:
\begeq
R^1_1(\theta)=\tanh \left({\theta\over 2}-{i\pi\over 4}\right).
\label{btiers}
\endeq
From the pole of the 2 solitons form factor, the one breather
form factor is found using (\ref{polerel}) and its contribution to the
conductance
is~:
\begeq
\Delta G(\omega)^{(1)}=-\mu^2 {\pi d^2\over 8} {\cal R}e \
\tanh\left[{\log({\omega\over \sqrt{2} T_B})\over 2}-i\pi/4\right],
\label{onebr}
\endeq
here $\mu=\pi$ is fixed by (\ref{valofmu}) and $d=0.1414...$.
The contribution from the two solitons form factors is computed
similarly,
we find~:
\begeqar
\Delta G(\omega)^{(2)}=&-{\mu^2 d^2\over 2} {\cal R}e
\int_{-\infty}^0~ d\theta~
{R(\theta+\log({\omega\over T_B})) R(\log [(1-e^\theta){\omega\over
T_B}]\over
\cosh^2(\theta-\log(1-e^\theta))}~ \vert
\zeta[\theta-\log(1-e^\theta)]\vert^2\nonumber\\
 &e^\theta \left[ e^\theta (1-e^\theta) ({\omega\over T_B})^2+
{1\over e^\theta (1-e^\theta) ({\omega\over T_B})^2}\right],
\label{deuxsol}
\endeqar
where $\zeta(\theta)$ is the function defined in (\ref{ffsmir}).
We can similarly write the following contribution, and we find that
these last
two expressions are sufficient for any reasonable purpose, they give
the
frequency dependent conductance to more than one percent accuracy.
We give the full function $G(\omega)$ in figure 3.

\begin{figure}[tbh]
\centerline{\psfig{figure=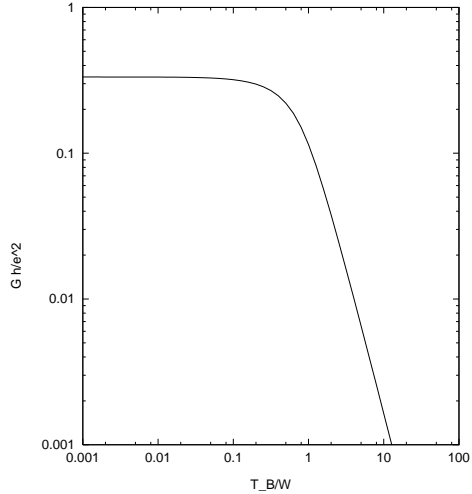,height=2.5in,width=2.5in}}
\caption{Frequency dependent conductance at T=0.}
\end{figure}

Observe that in
the UV and in the IR we obtain the $\omega$ dependance discussed
previously, even with the truncation to a few form-factors. The
form-factors
expansion is indeed  very different from the perturbative expansion
in powers of the
coupling constant in the UV, or in powers of the inverse coupling
constant
in the IR. Each form-factor contribution has by itself the same
analytical structure as the whole sum; contributions with higher
number of particles simply determine coefficients to a greater
accuracy.

\section{Anisotropic Kondo model and dissipative quantum mechanics.}

It turns out that the anisotropic Kondo model is related with 
the famous two state problem of  dissipative quantum mechanics
\cite{Leggett}.  The bosonised form of the
hamiltonian is~:
\begeq
H={1\over 2}\int_{-\infty}^0 dx
\left[8\pi g\Pi^2+{1\over 8\pi g}(\partial_x\Phi)^2
\right]+{\lambda\over
2}\left(S_+e^{i\Phi(0)/2}+
S_-e^{-i\Phi(0)/2}\right),
\label{hamil}
\endeq
where $S$ are Pauli matrices.  As for the quantum Hall problem, we
keep using
as a basis the massless excitations of the sine-Gordon model; however
the boundary
interation is different: this will result in different reflection
matrices.

\subsection{A computation of $C(t)$.}

We first work in imaginary time. We consider therefore
the anisotropic Kondo problem at temperature $T$.
Let us consider the quantity
$X(y)\equiv<[S^z(y)-S^z(0)]^2>$. On the one hand, using that
 $S^z=\pm 1$,
it reads $2[1-C(y)]$, where $C(y)$ is the usual spin correlation
\begeq
C(y)={1\over 2}\left[<S^z(y)S^z(0)>+
<S^z(0)S^z(y)>\right].
\label{corr}
\endeq
On the
other hand, we can write a perturbative expansion for $X(y)$ by
expanding
evolution operators  in powers of the coupling constant $\lambda$. At
every
 order,
we get ordered monomials which are are a product of a monomial
in $S^{\pm}$ and vertex operators of charge $\pm 1/2$. We must then
evaluate $S^z(y)-S^z(0)$ for each such term, trace over the
two possible spin states, and average over the quantum field. Since
we deal
 with  spin $1/2$, terms $S^+$ and $S^-$ must
alternate, and there must be an overall equal number of
$S^+$ and $S^-$, and an equal number of  $1/2$ and $-1/2$ electric
charges.

Now, since each $S^+(y)$ comes with a $e^{-i\Phi(y)/2}$
and each $S^-(y)$ comes with a $e^{i\Phi(y)/2}$, $S^z(y)=S^z(0)$
if there is a vanishing electric charge inserted between $0$ and $y$,
and $S^z(y)=-S^z(0)$ if the charge inserted between $0$ and $y$
is non zero (and then it has to be $\pm 1/2$). Therefore, we can
write the
perturbation expansion of $X(y)$ in such a way that
the spin contributions all disappear:
\begeqar
X(y)&=&{1\over Z}\sum_{n=0}^\infty \lambda^{2n}
\sum_{alternating ~\epsilon_i
=\pm}
\ \sum_{p=0}^{2n} \int_0^y dy_1\int_{y_1}^y
dy_2\ldots\int_{y_{p-1}}^y dy_p
\int_y^{1/T}dy_{p+1}\ldots \int_{y_{2n-1}}^{1/T}dy_{2n}\nonumber\\
&\times& 4(\epsilon_1+\ldots+\epsilon_p)^2\left<
e^{-i\epsilon_1\phi(y_1)/2}\ldots
e^{-i\epsilon_{2n}\phi(y_{2n})/2}\right>_N,
\label{main}
\endeqar
where $Z$ is the partition function, the factor 4 occurs
because of the normalization $S^z=\pm 1$, for every configuration of
$\epsilon$'s, only one value of $S^z(0)$ gives a non vanishing
contribution. Here, the label $N$ indicates
correlation functions for the free boson evaluated with Neumann
boundary conditions (the conditions as $\lambda\to 0$).

On the other hand,  let us consider
the correlator
\begeq
<\partial_x\Phi(x,y)\Phi(0,y')>_N
=-8g{x\over x^2+(y-y')^2},
\label{tatai}
\endeq
which goes to $-8g\pi\delta(y-y')$ as $x\to 0$. We have then, by
Wick's
theorem,
\begeqar
<e^{-i\epsilon_1\Phi(y_1)/2}\ldots 
e^{-i\epsilon_{2n}
\Phi(y_{2n})/2}
\partial_x\Phi(x,y)>_N=\nonumber\\ 
8ig\left(\sum_{i=1}^{2n} \epsilon_i{x\over x^2+(y-y_i)^2}\right)
\left<e^{i\epsilon_1\Phi(y_1)/2}\ldots e^
{i\epsilon_{2n}\Phi(y_{2n}/2)}\right>_N,
\label{tataii}
\endeqar
and therefore
\begeqar
\left<e^{-i\epsilon_1\Phi(y_1)/2
}\ldots e^{-i\epsilon_{2n}\Phi
(y_{2n})/2} :\partial_x\Phi(x,y)
\partial_x\Phi(x,y'):\right>_N=\nonumber\\
-(8g)^2\left(\sum_{i=1}^{2n} \epsilon_i{x\over x^2+(y-y_i)^2}\right)
\left(\sum_{i=1}^{2n} \epsilon_i{x\over x^2+(y'-y_i)^2}\right)
\left<e^{-i\epsilon_1\Phi(y_1)/2}\ldots
 e^{-i\epsilon_{2n}\Phi(y_{2n})/2}\right>_N,
\label{tataiii}
\endeqar
where contractions between the dots $:$ are discarded. In (\ref{tataiii}),
contractions between the dots would lead to a term factored out as
the
product of the two point function of $\partial_x\Phi$  and the $2n$
point
function of vertex operators, both evaluated with N boundary
conditions. Now, we are going to
be interested in the $x\to 0$ limit where, with N boundary
conditions, $\partial_x\Phi$ vanishes.   As a result we can actually
forget the subtraction in \ref{tataiii}, and write simply
obtain
\begeq
X(y,\lambda)=-{1\over (4g\pi)^2}\lim_{x\to 0}
\int_0^y\int_0^y dy'dy''<\partial_x\Phi(x,y')\partial_x
\Phi(x,y'')>_\lambda,
\label{maini}
\endeq
where the label $\lambda$ designates the correlator
evaluated at coupling $\lambda$, N corresponding to $\lambda=0$
. Hence, we can get $C(y)$ from the current current correlator. The
latter
can then be obtained using form factors
along the above lines. The only difference is the boundary
matrix.
 If we restrict to the repulsive regime where the bulk spectrum
contains only a soliton and an antisoliton,
 one has~:
\begeq
 R_\pm^\mp=\tanh\left({\theta\over 2}-{i\pi\over 4}\right), \ \
R^\pm_\pm=0.
\label{bsol}
\endeq
Here again our conventions are such that
a soliton bounces back as an antisoliton, in agreement
with the    UV {\bf and}
the IR limit that have Neumann boundary conditions. In the attractive
regime we need to add the breathers with~:
\begeq
R^m_m={\tanh({\theta\over 2}-{i\pi m\over 4 (1/g-1)})\over
\tanh({\theta\over 2}+{i\pi m\over 4 (1/g-1)})}.
\label{bbreath}
\endeq
Writing, as in (\ref{idenmaktata})~:
\begeq
<\partial_{\bar{z}}\Phi(x,y')
\partial_{z}\Phi(x,y'')>_{\lambda}
=\int_0^\infty {\cal G}(E,\beta_B)\exp\left[2Ex-iE(y'-y'')\right],
\label{titi}
\endeq
we have that~:
\begeqar
&\lim_{x\to 0}<\partial_x\Phi(x,y')\partial_x
\Phi(x,y'')>_\lambda=\nonumber\\
&\int_0^\infty dE\left[{\cal G}(E,\beta_B)-
{\cal G}(E,-\infty)\right]\exp[-iE(y'-y'')]+c.c. ,
\label{titii}
\endeqar
where the
 $<\partial_z\Phi\partial_
z\Phi>$ part and its complex conjugate (which are $\lambda$
independent) have been evaluated by requiring that
the correlator vanishes as $\lambda\to 0$ due to $N$ boundary
conditions. Hence, using the fact that ${\cal G}$ is real,
\begeq
X(y)={1\over 2(g\pi)^2}\int_0^\infty {dE\over
E^2}\left[{\cal G}
(E,\theta_B)-
{\cal G}(E,-\infty)\right]\sin^2 (Ey/2).
\label{super}
\endeq
Therefore, if we write~:
\begeq
C(y)-1=\int_0^\infty A(\omega_M) \cos(\omega_M y) \
d\omega_M,
\label{fou}
\endeq
where $\omega_M$ is a Matsubara frequency, we have~:
\begeq
A(\omega_M)={1\over
(2g\pi)^2}{1\over\omega_M^2}\left[{\cal G}
(\omega_M,\theta_B)-
{\cal G}(\omega_M,-\infty)\right].
\label{mainfou}
\endeq

An observation is now in order. From the foregoing results
we see that
\begeq
<S^z(0)S^z(y)>-1={1\over (2g\pi)^2}\int_0^\infty {dE\over
E^2}
\left[{\cal G}
(E,\theta_B)-
{\cal G}(E,-\infty)\right]\cos Ey.
\label{adji}
\endeq
On the other hand, consider the expression
\begeq
<\int_{-\infty}^0 dx'\int_{-\infty}^0 dx''
\left[<\partial_x\Phi(x',y)
\partial_x\Phi(x'',0)>_\lambda-<\partial_x\Phi(x',y)
\partial_x\Phi(x'',0)>_N\right].
\label{adjii}
\endeq
By using the same representation (\ref{titi}), this is
$$
\int_{-\infty}^0 dx'\int_{-\infty}^0 dx'' \int_0^\infty dE
\left[{\cal G}
(E,\theta_B)-
{\cal G}(E,-\infty)\right] \exp\left[E(x'+x'')-iEy\right]+cc,
$$
which coincides with (\ref{adji}) after performing the integrations. We
conclude
that
\begeq
<S^z(0)S^z(y)>-1=<{\cal J}_x(0){\cal J}_x(y)>-
<{\cal J}_x(0){\cal J}_x(y)>_N,
\label{surprise}
\endeq
where we defined
\begeq
{\cal J}_x={1\over 2g\pi}\int_{-\infty}^0
\partial_x\Phi(x,y) dx.
\label{deffij}
\endeq
We find also by the same manipulations that
\begeq
<S^z(0)S^z(y)>-1=<{\cal J}_y(0){\cal J}_y(y)>-<{\cal
J}_y(0){\cal J}_y(y)>_N,
\label{surprisei}
\endeq
where
\begeq
{\cal J}_y={1\over 2g\pi}\int_{-\infty}^0
\partial_y\phi(x,y) dx.
\label{deffiij}
\endeq

We now continue to real frequencies to find the response function~:
\begeq
\chi''(\omega)\equiv {1\over 2}\int dt e^{i\omega t}\left<
[S^z(t),S^z(0)]\right>,
\label{resp}
\endeq
to find~:
\begeq
\chi''(\omega)={1\over (2g\pi)^2}{1\over\omega^2}
\hbox{Im} \left[{\cal G}
(-i\omega,\theta_B)-
{\cal G}(-i\omega,-\infty)\right].
\label{mainmain}
\endeq

As a first example, let us consider the so called Toulouse limit
or  free fermion case. Then the only contribution comes from the
soliton
antisoliton
form factors, which as discussed above is $f(\theta_1,\theta_2)=i\mu 
e^{\theta_1/2} e^{\theta_2/2}$. Hence,
\begeq
\chi''(\omega)={1\over \pi^2} \hbox {Re}
\int_{-\infty}^\infty
\int_{-\infty}^\infty d\theta_1 d\theta_2
{e^{\theta_1}e^{\theta_2}\over (e^{\theta_1}+ie^{\theta_B})
(e^{\theta_2}+ie^{\theta_B})}
{1\over
e^{\theta_1}+e^{\theta_2}}\delta(e^{\theta_1}+e^{\theta_2}-\omega),
\label{freemain}
\endeq
that is
\begeq
\chi''(\omega)={2\over \pi^2}{T_B\over\omega}\hbox
{Im}
 \left(
\int_0^\omega dx{1\over  (x+iT_B)(\omega-x+iT_B)}\right),
\label{freemaini}
\endeq
and
\begeqar
\chi''(\omega)&=&{4\over \pi^2}{T_B\over\omega}
\hbox {Im}{1\over \omega+2iT_B}
 \ln\left({\omega+iT_B\over iT_B}\right)\nonumber\\
&=&{1\over \pi^2}
{4T_B^2\over  \omega^2 +4T_B^2}\left[{1\over\omega}
\ln\left({T_B^2+\omega^2\over T_B^2}\right)+{1\over T_B}
\tan^{-1}{\omega\over T_B}\right].
\label{resu}
\endeqar

In general, observe that, since the reflection matrix for solitons
and antisolitons expands as a series in $e^\beta$,
${\chi''(\omega)\over\omega}$, will,
for any coupling, expand as a series of the form $(\omega/T_B)^{2n}$
in the
IR. In particular, this leads to
a behaviour $C(t)\propto {1\over t^2},t>>1$ for any $g$. In the UV,
one has to
split integrals in two pieces. Since
the
soliton-soliton form factors expansion involves powers of $\exp
({1\over
g}-1)\theta$, ${\chi''(\omega)\over\omega}$ expands as a double series
in
$(T_B/\omega)^{2-2g}$ and $(T_B/\omega)^2$ in the UV. Hence at short
times,
$C(t)-1\propto t^{2-2g}$. This is in agreement with the qualitative
analysis of \cite{Guinea}.

Results for $g\neq 1/2$ are more involved because there are non zero
form factors at all levels.  Still, when working out the first few
form
factors we observe a very rapid convergence with the number of
rapidities and again we can give precise results for different
values of $g$.  As an example, let us show the results for
$g=1/3$.

The computation for $g=1/3$ is very similar to the previous
conductance
computations.  The boundary matrices are much more simpler though.
In this case we have~:
\begeq
R_\pm^\mp=\tanh \left({\theta\over 2}-{i\pi\over 4}\right),\ \ R_\pm^\pm=0,
\label{konsol}
\endeq
and~:
\begeq
R_1^1={\tanh({\theta\over 2}-{i\pi\over 8})\over
\tanh({\theta\over 2}+{i\pi\over 8})}.
\label{konbrea}
\endeq
Then, as was found for the conductance, we find that the first two
contributions are sufficient for most purposes, they are given
by~:
\begeq
\Delta\chi''(\omega)^{(1)}=-{9 \mu^2  d^2\over 8\pi  \omega }\hbox{Re }
\left[ {\tanh\left({\log({\omega\over \sqrt{2}T_B})\over
2}-{i\pi\over 8}\right)
\over
\tanh\left({\log({\omega\over \sqrt{2}T_B})\over 2}+{i\pi\over
8}\right)}
-1\right]
\label{premcontrib}
\endeq
and~:
\begeqar
\Delta\chi''(\omega)^{(2)}=&-&\left( {3 \mu d\over 2 \pi}\right)^2
{1\over \omega} \hbox{Re }
\int_{-\infty}^0 d\theta {\vert \zeta(\theta-\log
(1-e^\theta))\vert^2\over
\cosh^2(\theta-\log (1-e^\theta))} e^\theta \nonumber\\
&\times& \left[ R^+_- (\theta+\log(\omega/T_B))
R^+_-(\log[(1-e^\theta)\omega/T_B])-1\right].
\label{seccontrib}
\endeqar
Again these two expressions are sufficient to get a very precise
result.
Similar computations give rise to the results in
figure 4 where we plotted  $S(\omega)\equiv\chi''(\omega)/\omega$  for the values,
$g=3/5,1/2, 1/3, 1/4$.

\begin{figure}[tbh]
\centerline{\psfig{figure=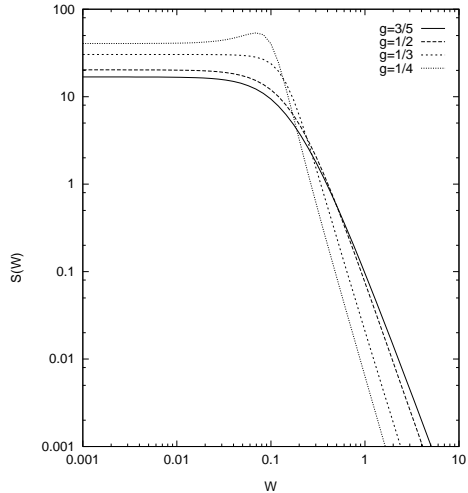,height=2.5in,width=2.5in}}
\caption{Spectral function for $T_B=0.1$.}
\end{figure}

When making these calculations we have to be careful about which
terms are
needed for a good convergence.
 Our observation is that keeping the form
factors up to two rapidities give very good results precise to $1\%$.
 It is
possible to go further and get a better precision if needed.  The
last
statements are true for $g\in [0.6,0.2]$ and we believe even further
(we can get rough bounds on the higher contributions and have an
idea of the precision).  Still, for the moment the isotropic Kondo
point is difficult to treat.

To a good approximation, $S(\omega)$ can be represented for $g< {1\over 2}$ by a Lorentzian shape built around the 
{\sl four} poles of the one-breather term:
\begeq
S(\omega)\approx 
{cst\over \left[\left({\omega\over \Omega}\right)^2-1+\left({\Gamma\over\Omega}\right)^2\right]^2
+4\left({\Gamma\over\Omega}\right)^2} \label{lorentz}
\endeq
where the poles are located at $\omega=\pm \Omega+\pm i\Gamma$. Here, it follows from the one breather reflection matrix that
\begeq
{\Omega\over \Gamma}=\cot {\pi g\over 2(1-g)},
\endeq
and $\Omega=T_B \sin {\pi g\over (1-g)}$. The approximation improves as $g$ gets smaller.

According to (\ref{lorentz}),
the pic in $S(\omega)$ disappears at $g={1\over 3}$, where $\Omega=\Gamma$. The poles meanwhile reach
the imaginary axis at $g={1\over 2}$ where ${\Omega \over \Gamma}=0$. These features are shared by the 
exact solution, as far as can be established from numerical study of the form-factors series: the pic
in $S(\omega)$ disappears at $g={1\over 3}$, while the tail oscillations of $C(t)$ disappear at $g={1\over 2}$. 
Depending on one's definition of the ``transition'' from coherent to incoherent regime, the latter therefore
occurs at $g={1\over 3}$ or $g={1\over 2}$. This feature was somewhat unexpected 
from previous approximate  studies, but is fully confirmed by recent numerical 
work \cite{Costi,Volker}. In any case, there is no real transtion per se, 
and it is possible that the study of different quantities (for instance multi point correlations)
would exhibit a qualitative change of behaviour at still another value of $g$.

\subsection{Shiba's Relation.}

Up untill now, we showed results for certain values of $g$ more or
less limited by our ability (or tenacity) to write the form factors
corresponding to that value of the anisotropy, and make them
converge.  It is not impossible to find general relations though;
for  example the behaviours in the UV and the IR in different models
were infered in all generality.

Here we present a generalisation of Shiba's relation \cite{Shiba} which was proven
for the Anderson model and generalised to Luttinger liquids
by Sassetti and Weiss \cite{Shigen}. The relation states that~:
\begeq
\lim_{\omega\rightarrow 0} {\chi''(\omega)\over \omega}
=2 \pi g\chi_0^2
\label{shib}
\endeq
with $\chi_0$ the static succeptibility.    If we look at the
quantity~:
\begeqar
{\cal  G}(E)&=&E
 \sum_{n=0}^\infty \int_{-\infty}^0 {d\theta_1\ldots d\theta_{2n-1}
\over(2\pi)^{2n}(2n)!}{1\over 1-e^{\theta_1}-\ldots
-e^{\theta_{2n-1}}}\nonumber\\ &\times&
K^{a_1b_1}(\ln (T_B/E)-\theta_1)\ldots K^{a_{n-1}b_{2n-1}}(\ln
(T_B/E)-\theta_{n-1})
\nonumber\\
&\times&K^{a_{2n}b_{2n}}\left[\ln
(T_B/E)-\ln\left(1-e^{\theta_1}-\ldots-e^{\theta_{2n-1}}\right)\right]\nonumber\\
&\times&f_{a_1\ldots a_{2n}}f^*_{b_1\ldots b_{2n}}\left[\theta_1\ldots\theta_{2n-1},
\ln\left(1-e^{\theta_1}-\ldots-e^{\theta_{2n-1}}
\right)\right],
\label{cenicetata}
\endeqar
insert it in the expression for $\chi''(\omega)$
and expand it around $E\simeq 0$ we find that the contributions from
the $K$ matrices all cancel and only a constant is left (we have
to take into account the fact that the soliton/anti-solitons
$K$ matrices always appear in pair).
Then comparing
this with the UV normalisation we find that~:
\begeq
\lim_{\omega\rightarrow 0} {\chi''(\omega)\over \omega}=
{1\over \pi^2 g T_B^2}.
\label{omegzero}
\endeq
The total succeptibility is $\chi=\chi'+i \chi''$ and
the static succeptibility $\chi_0$ which is the
zero frequency limit of $\chi'$ can also be infered from
the previous expressions for the spin-spin correlation.
We just need to take the real part when continuing (\ref{mainfou})
to real frequencies, which leads to~:
\begeq
\chi_0={1\over \pi^2} g T_B.
\label{statsuccep}
\endeq
Finally, in order to make contact with the usual form of Shiba's relation,
we need to renormalise the spins to $1/2$ and use the usual
normalization for Fourier transforms, which leads indeed to (\ref{shib}). 

\section{Friedel oscillations: correlations involving Vertex operators}

As exemplified in the previous sections, 
the method works naively indeed   for currents, ie
for operators with no anomalous dimension (calculations 
for the stress energy tensor for instance, would be very similar). Many
physical properties
are however described by more complicated operators, that is 
operators which have a 
non trivial anomalous 
dimension.
As an example, I would like to discuss here the equivalent of Friedel oscillations in
Luttinger liquids: more precisely, 
the  $2k_F$ part of the
charge density profile in a one dimensional Luttinger liquid
away from an  impurity. This is  a problem which has attracted a fair amount of interest 
 recently \cite{Grabert},\cite{Schmitteckert}
.

We start with the bosonised form of the model. The
hamiltonian is:
\begin{equation}
H={1\over 2}\int_{-\infty}^\infty dx \ [8\pi g\Pi^2+\frac{1}{8\pi g}
(\partial_x \phi)^2]+
\lambda \cos\phi(0),
\end{equation}
where we have set $v_F=g$.  Then for the  Friedel oscillations,
the charge density operator is just~:
\begin{equation}
\rho(x)=\rho_0+2 \partial_x\phi +
\frac {k_F}{\pi} \cos[2k_Fx+\phi(x)].
\end{equation}
with $\rho_0=\frac{k_F}{\pi}$ the background charge. 

We
decompose  this system into even and odd basis (this is was explained already in \cite{Houches98})
 by
writing $\phi=\phi_L+\phi_R$ and setting~:
\begin{eqnarray}
\varphi^e(x+t)={1\over\sqrt{2}}\left[\phi_L(x,t)+\phi_R(-x,t)\right]
\nonumber\\
\varphi^o(x+t)={1\over\sqrt{2}}\left[\phi_L(x,t)-\phi_R(-x,t)\right]
\end{eqnarray}
Observe that these two field are left movers. We now fold the system
by setting~ (prefactors here are slightly 
different from (\cite{Houches98}) because of the normalizations
in the hamiltonian)~:
\begin{eqnarray}
&\phi^e_L=\sqrt{2}\varphi^e(x+t),\ x<0\ \ \ \ 
\phi^e_R=\sqrt{2}\phi^e(-x+t),
x<0\nonumber\\
&\phi^o_L=\sqrt{2}\varphi^e(x+t),\ x<0\ \ \ \ 
\phi^o_R=-\sqrt{2}\phi^e(-x+t), x<0
\end{eqnarray}
and introduce new fields $\phi^{e,o}=\phi^{e,o}_L+\phi^{e,o}_R$.
The density oscillations now read~:
\begin{equation}
\frac{\langle\rho(x)-\rho_0\rangle}{\rho_0}=\cos(2k_Fx+\eta_F)
\langle \cos\frac{\phi^o(x)}{2}\rangle \langle
\cos\frac{\phi^e(x)}{2}
\rangle,
\end{equation}
with $\eta_F$ the additional phase shift coming from the unitary
transformation to eliminate the forward scattering term \cite{Grabert}.
$\phi^o$ is the odd field with Dirichlet boundary conditions,
at the origin $\phi^o(0)=0$ leading to \cite{Fabrizio}~:
\begin{equation}
\langle \cos\frac{\phi^o}{2}\rangle \propto
\left(\frac{1}{x}\right)^{g/2},
\label{diri}
\end{equation}
and the $\phi^e$ part is computed with the hamiltonian~:
\begin{equation}
\label{hamili}
H^e=\frac{1}{2}\int_{-\infty}^0 dx \
[8\pi g\Pi^{e2}+\frac{1}{8\pi g}(\partial_x\phi^e)^2]+
\lambda \cos\frac{\phi^e(0)}{2}.
\end{equation}
On general grounds, we expect the scaling form~:
\begin{equation}
\langle \cos\frac{\phi^e}{2}\rangle \propto
\left(\frac{1}{x}\right)^{g/2}F(\lambda x^{1-g}),
\label{scalingfct}
\end{equation}
where $F$ is a scaling function to be determined. Note that
even the small $x$ behaviour of this function was not known in
general.

To compute the  correlation functions , we use the formalism introduced in the previous section (with $\phi^e\equiv\Phi$).
The massless scattering description and the boundary state are the same; only the operator which is studied 
changes.

The one point function of interest is
$\langle 0|\cos\frac{\Phi}{2}|B\rangle$. 
We thus need   the matrix elements of the operator
$\cos\frac{\Phi}{2}$
in the quasiparticle basis: these still 
follow easily from the massive sine-Gordon form-factors
 \cite{Smirnov}, like the form-factors of $\partial\Phi$ in the previous sections. Difficulties however arise  when one considers
 the contribution of each form-factor to the one-point function: the rapidity 
integrals turn out to be all IR divergent! This was
not the case for  the current operator, whose form factor
has the naive engineering dimension of an energy, leading to
convergent integrals: some sort of additional regularization is  thus needed 
here. 

To explain the strategy,  consider first
the case  $g=1/2$. Here,
the friedel oscillations are simply \cite{LLS}
related to the   spin one point function in an
Ising model with  boundary magnetic field. By using
the same approach as the one outlined before,
one finds the following form-factors expansion~:
\begin{eqnarray}
\langle \sigma(x)\rangle&=&\sum_{n=0}^\infty \frac{1}{n!}
\int_{-\infty}^\infty \prod_{i=1}^n \left\{
\frac{d\theta_i}{2\pi} \tanh\frac{\theta_B-\theta_i}{2}
e^{-2m x e^{\theta_i}}\right\} \nonumber \\
&\times&\prod_{i<j} \left(\tanh\frac{\theta_i-\theta_j}{2}\right)^2.
\end{eqnarray}
The integrals are all divergent at low energies,
when $\theta_i\rightarrow
-\infty$ and  the integrand goes to a constant. Let
 us  then introduce an IR cut-off ( we chose
 $\theta\geq \theta_{min}$ and set $\Lambda\equiv e^{\theta_{min}}$)
and
take the log of the previous expressions (a similar method has been
used in \cite{Smirnov} to study the UV limit
of massive correlators. See also \cite{Yurov},\cite{MussardoCardy}).
 Ordering this log by
increasing
number of integrations, one can show that each term
diverges as $\ln\Lambda$. Moreover, since
the divergence occurs at very low energy, where the $\tanh$ goes to
unity,
the amplitudes of these $\ln\Lambda$ do {\bf not} depend
on $\theta_B$ (for $\theta_B\neq -\infty$), ie on the boundary
coupling. It is
then easy to get rid of the cut-off: we simply
substract
the log of the IR spin function, ie we substract the same formal
expression with $\theta_B= \infty$. The first two terms of the
resulting expression read~:
\begin{eqnarray}
\ln\frac{\langle \sigma(x)\rangle_{T_B}}{\langle
\sigma(x)\rangle_{IR}}
&=&\int_{\Lambda}^\infty \frac{du}{2\pi u} e^{-2u x}
\left( \frac{T_B-u}{T_B+u}-1\right)\nonumber \\
&+&\frac{1}{2}\int_\Lambda^\infty \prod_{i=1}^2 \frac{du_i}{2\pi u_i}
e^{-2\mu  u_i x} \left( \prod_{i=1}^2
\frac{T_B-u_i}{T_B+u_i}-1\right)
\nonumber \\ &\times&
\left[\left(\frac{u_1-u_2}{u_1+u_2}\right)^2-1\right]+\cdots
\label{sigff}
\end{eqnarray}
where we have set  $\mu=1$, $u_i=e^{\theta_i}$,
$T_B=e^{\theta_B}\propto \lambda^{1/(1-g)}$.

\begin{figure}[tbh]
\centerline{\psfig{figure=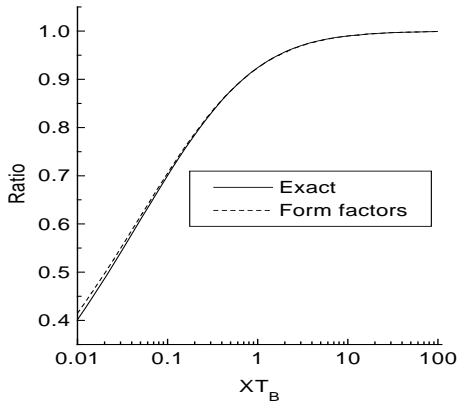,height=2.5in,width=2.5in}}
\caption{Accuracy of the finite $T_B$ over the IR
part of the envelope of $\rho(x)$ for $g=1/2$.}
\end{figure}

Clearly, the integrals are now convergent at low energies, and we can
send $\Lambda$
to zero. Since the IR value of the one point function is easily
determined by other means, $<\sigma(x)>_{IR}\propto x^{-1/8}$
\cite{CardyLewellen},
we can now obtain $<\sigma(x)>_{T_B}$ from (\ref{sigff}). Hence the
procedure involves a {\sl double} regularization. Of course,
there
remains an infinity of terms
to sum over. However, as in the case of current operators, the
convergence
of the form-factors expansion is very quick, and the first few terms
are sufficient to get excellent accuracy all the way from UV to
IR. To illustrate this more precisely, we recall
that  for $g=1/2$ (\ref{sigff}) can be resummed in closed form \cite{Chatterjee,LLS},
giving rise to~:
\begin{equation}
R_{exact}=
\frac{\langle \sigma(x)\rangle_{T_B}}{\langle \sigma(x)\rangle_{IR}}
=\frac{1}{\sqrt{\pi}} \sqrt{2 xT_B} e^{xT_B} K_0(x T_B).
\label{exact}
\end{equation}
By reexponentianing  the two first terms in (\ref{sigff}),
one gets a ratio differing from (\ref{exact})
by at most $1/100$ for $xT_B\in [0,\infty)$ (see figure 1).

By reexponentiating
the first three terms, acuracy is improved to more than $1/1000$.
Clearly,
the form-factors approach thus provides  analytical expressions
that can be considered as exact for most reasonable purposes.

It is fair to mention however that, at any given order in
(\ref{sigff}), the
exponent controlling the $x\to 0$ behaviour is not exactly
reproduced, as could be seen on a log-log plot.
For instance, the first term is immediately found to produce a
behaviour $R(x)\propto x^{1/\pi}$,
to be compared with the result $R_{exact}(x)\propto x^{1/2}\ln x$.
The
comparison  of the exact result (\ref{exact}) and of (\ref{sigff})
show that the form factors expression has, term by term, the correct
asymptotic expansion
ie the  IR expansion in powers of ${1\over xT_B}$. Adding terms with
more form-factors simply gives a more accurate determination
of the coefficients. This is to be compared with   the
results of \cite{LSS} for eg the frequency dependent conductance,
where the form factors expression had the correct functional
dependence both in the UV and in the IR.  This is not to say that the
method
is inefficient in the UV, because we know, at least formally, all the
terms.
In fact, we will show in what follows how the expansion (\ref{sigff}) can always be
resummed
in the UV, and that the exponent can be exactly obtained from the form-factors
approach too.

The regularization  is the
same for other values of $g$. Let us discuss here 
 the cases $g=1/t$ with $t$ integer again.  For these values,
the scattering  is diagonal and
the form factors  are rather simple.  To obtain
them, we again take the  massless limit of the results
in \cite{Smirnov} (but this time for vertex operator $\cos\Phi/2$ (in normalizations where the bulk sine-Gordon perturbation is $\cos\Phi$) instead
of the current)
and impose  that half of the
quasiparticles  become right movers and half become left movers,
since
the boundary state always involve pairs of right and left moving
particles.  It is in fact easier  to take that limit if we change
basis from the solitons and anti-solitons to $\frac{1}{\sqrt{2}}(
|S>\pm |A>)$.  In that case, the boundary scattering matrix
becomes diagonal and the isotopic indices always come in pairs.
The reflection matrices in this new basis are given by~:
\begin{eqnarray}
&R_-(\theta)=-e^{i\frac{\pi}{4}(2-t)} \tanh\left[\frac{(t-1)\theta}{2}
+i\frac{\pi(t-2 )}{4 }\right] R(i\frac{\pi}{2}-
\theta) \nonumber \\
&R_+(\theta)=e^{i\frac{\pi}{4}(2-t)} R(i\frac{\pi}{2}-\theta)
\end{eqnarray}
with (note the slight change of notation compared with (\ref{norm}))~:
\begin{equation}
R(\theta)=\exp\left( i \int_{-\infty}^\infty \frac{dy}{2y}
\sin\frac{2(t-1) y\theta}{\pi}\frac{\sinh(t-2)y}
{\sinh(2y)
\cosh(t-1)y}\right).
\end{equation}
The breathers reflection matrices follow from  \cite{Ghoshal} (see also (\ref{btiers}).

The case $g=1/2$ having already been worked out,  let us concentrate
on
$g=1/3$.  There, in addition to the soliton and
anti-soliton,
there is also one breather.
The first contribution to the one point function comes  from
the two breathers  form-factor, with one right moving and
one left moving breather. It is given by a constant~:
\begin{equation}
f(\theta,\theta)_{11}^{LR}=c_1,
\end{equation}
and this obviously leads to IR divergences.
Other contributions come from $2n$ breathers form-factors, and $4n$
solitons form-factors
The whole expression can be controlled as for $g=1/2$,
by taking the log, and factoring out the IR part.  Setting
$c(x)= \cos\frac{\phi(x)}{2}$, we
organize the sum as follows~:
\begin{equation}\label{tata}
\ln\frac{\langle c(x)\rangle_{T_B}}{\langle c(x)\rangle_{IR}}=\ln
R^{(2)}+
\ln R^{(4)}+
\cdots
\end{equation}
with the subscript denoting  the number of intermediate excitations.

Then, using the explicit expressions for $g=1/3$ we find~:
\begin{equation}
\ln R^{(2)}=2 c_1 e^{2\sqrt{2}T_B x} Ei(-2\sqrt{2}T_B x),
\end{equation}
where $E_i$ is the standard exponential integral. The next
term $\ln R^{(4)}$ is a bit bulky to be written here,
but it is very easy to obtain - similar expressions
have been explicitely given in the previous sections.This
is all what is needed for an accuracy better than 1 percent.
In figure 6 we present the results of the ratio at
$g=1/2,1/3,1/4$ for the Friedel oscillations.
It should be noted that this ratio is just the pinning function
of reference \cite{Grabert} and our results agree well qualitatively
with the results found there.

As mentioned before, the deep UV behaviour is a little more difficult
to obtain: the accuracy is good because the ratio goes to zero
anyway,
but the numerical evaluation of the power law would not be too
accurate
with the number of terms we consider. Fortunately,
the full form-factors expansion allows the analytic determination
of this exponent. First, observe
for instance that in (\ref{sigff}) the integrals converge for all
$T_B\neq 0$,
but strictly at $T_B=0$, they do not. To get the dependence of
$\langle c(x)\rangle$
as $T_B\to 0$, we will consider, the logarithm of another ratio,
$\ln\frac{\langle
c(x)\rangle_{T_B}}{\langle c(x')\rangle_{T_B}}$,
where $x$ and $x'$ are two arbitrary coordinates. For this ratio,
even at $T_B=0$,
the integrals are convergent. But $T_B=0$ is the UV fixed point, with
Neumann boundary conditions. While the one point function $\langle
c(x)\rangle_{UV}$
vanishes, the ratio of two such one point functions is well defined,
and can be computed by putting an IR cut-off (a finite system). One
finds
that it goes as $(x/x')^{g/2}$. By regularity as $T_B\to 0$, the same
is true
for the ratio close to $T_B=0$, and thus one has
\begin{equation}
\langle c(x)\rangle \propto (xT_B)^{g/2}, x (T_B)\to 0
\end{equation}
This shows that the universal scaling function in (\ref{scalingfct})
behaves as
$F(y)\propto y^{g\over 1-g}$ for $g<{1\over 2}$. This exponent can
actually be obtained by perturbation theory. Indeed, the first term
in the perturbative expansion of  $\langle c(x)\rangle$ is
\begin{equation}
\lambda x^{g/2}\int_{-\infty}^\infty {dy\over (x^2+y^2)^g}
\end{equation}
For $g<1/2$, this integral diverges in the IR. To regulate it, we
need
to put a new cut-off: since there is no other length scale in the
problem, this
can be nothing but $1/T_B$. Changing variables, the leading behaviour
is $x^{g/2}T_B^g\propto x^{g/2}\lambda^{g/1-g}$, in agreement
with the previous discussion. The exponent   coincides
with the result of the self-consistent harmonic approximation
\cite{Grabert};
but it is important to stress that   the latter is valid only for
$g<<1$.

\begin{figure}[tbh]
\centerline{\psfig{figure=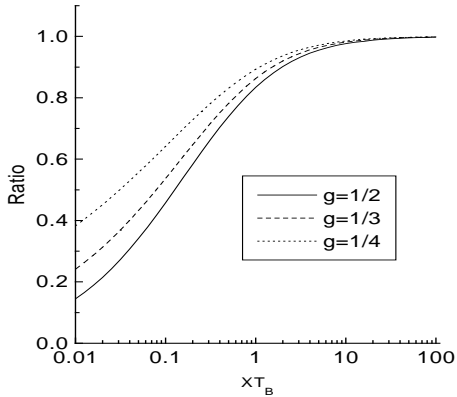,height=2.5in,width=2.5in}}
\caption{Ratio of the finite $T_B$ over the IR
part of the envelope of $\rho(x)$..}
\end{figure}

The function $F(y)$ behaves a $y \ln y$ for $g=1/2$. For
$g>1/2$, its
behaviour is simply $F(y)\propto y$, as can be easily shown
since the perturbative approach is now convergent.
As we approach $g=1/2$ this exponent seems to become asymptotic
and is more difficult to get numerically \cite{Grabert}.

\section{Conclusion}

The description of the Hilbert space in terms of the integrable 
quasiparticles basis allows, in a wide variety of cases,
extremely accurate computations of time and space dependent correlators. 
The  related form-factors method is not especially elegant, but gives rise 
to susprisingly good results: in fact, it is not clear whether exact analytical expressions
of the correlation functions - if ever obtained - will be more useful.
 Although we limited ourselves
to $g=1/t$ with $t$ integer in these notes,
all values of $g<1/2$ are accessible, but
general computations are more  complicated since the bulk scattering is non
diagonal.
The methods explained here  should be generalizable to other problems, in particular
the  determination of the screening cloud in the anisotropic Kondo
model \cite{Affleck}.

The region $g>1/2$ - in particular the $SU(2)$ symmetric point $g=1$ 
\footnote{Of course
the problem of Friedel oscillations for $g=1$ can be solved by fermionization \cite{Grabert}, since 
it corresponds to non interacting fermions.  In our
approach,
this point is non trivial
because of  the folding. This folding however is
necessary for any value $g\neq 1$: except at $g=1$, the problem on
the
whole line would not be integrable.}
 presents additional
difficulties, which are not yet solved. The situation does not look
desperate however.

As a last but important comment, I would like to stress that the 
form-factors technique has also been applied succesfully to the computation
of quantities of experimental interest in the case of (bulk) massive theories. 
See the lectures by F. Essler in this school, or the papers \cite{CM,BN,Essler}. One point functions in massive theories and in the presence of a boundary have also been recently studied in \cite{DPTW}.

\bigskip
\noindent{\bf Acknowledgments}: These notes result from close collaborations
with F. Lesage and S. Skorik, to whom I am very thankful. Support 
from the DOE, the NSF and the Packard Foundation is also gratefully acknowledged.


\begin{thebibliography}{999}



\bibitem{Houches98} H. Saleur, in Proceedings of the 1998 Les Houches Summer School,
``Topological Aspects of Low Dimensional Systems'', cond-mat/9812110

\bibitem{Karowski78} M. Karowski and P. Weisz, Nucl. Phys. B139 (1978) 445; B. Berg, M. Karowski and P. Weisz, Phys. Rev. D19 (1979) 2477; M. Karowski, Phys. Rep. 49 (1979) 229. 


\bibitem{Smirnov} F.A. Smirnov, ``Form factors in completely
integrable models of quantum field theory", World scientific
and references therein.

\bibitem{Mussardo94} G. Mussardo, ``Spectral representation of correlation functions
in two dimensional quantum field theories'', hep-th/9405128.

\bibitem{Others} V. Brzahnikov, S. Lukyanov, Nucl. Phys. B512 (1998) 616;
H. Babujian, A. Fring, M. Karowski, A. Zapletal, Nucl. Phys. B538 (1999) 535. 

\bibitem{ZamoZamo79} A. B. Zamolodchikov and Al. B. Zamolodchikov, Ann. Phys. 120 (1979) 253.

\bibitem{Korepin} V. E. Korepin, N.M Bogoliubov, A.G. Izergin, ``Quantum
inverse Scattering method and correlation functions", Cambridge
university press, 1993.



\bibitem{Mussardo92} G. Mussardo, Phys. Rep. 218 (1992) 215

\bibitem{Dorey98} P. Dorey, ``Exact S matrices'', hep-th/9810026

\bibitem{Kyoto} M. Jimbo and T. Miwa, ``Algebraic analysis of solvable lattice models'', AMS 1995, and references therein; N. Kitanine, J. M. Maillet and 
V. Terras, Nucl. Phys. B554 (1999) 647.

\bibitem{LSZ} H. Lehmann, K. Symanzik and W. Zimmermann, Nuovo Cimento 1 (1955) 205. 


\bibitem{Fring} A. Fring, G. Mussardo and P. Simonetti, Nucl. Phys. B393 (1993) 413.


\bibitem{Aldo} G. Delfino, G. Mussardo and P. Simonetti, Phys. Rev. D51 (1995) 6620.


\bibitem{Aldophd} G. Delfino, PhD Thesis, unpublished.

\bibitem{Cardy} J. Cardy, G. Mussardo, Nucl. Phys. 
B 410 (1993), 451.

\bibitem{GZ} S. Ghoshal, A.B. Zamolochikov, Int. J. Mod. Phys. A
 9, (1994) 3841.


\bibitem{KaneFisher} C.L. Kane, M.P.A. Fisher, Phys. Rev. {\bf B}46,
(1992) 15233.

\bibitem{Moon} K. Moon, H. Yi, C.L. Kane, S.M. Girvin, M.P.A. Fisher,
Phys. Rev. Lett. 71, (1993) 4391.

\bibitem{FLS} P. Fendley, A.W.W. Ludwig, H. Saleur, Phys. Rev. Lett.
74, (1995) 3005.

\bibitem{FSW} P. Fendley, H. Saleur, N.P. Warner, Nucl. Phys.
B 430 [FS], (1994) 577.

\bibitem{Wong} E. Wong and I. Affleck, Nucl. Phys. B417 (1994) 403. 

\bibitem{Guineaetal} F. Guinea, G. Gomez-Santos,
M. Sassetti, M. Ueda, Europhys. Lett. 30, 561 (1995)

\bibitem{Leggett} A.J. Leggett, S. Chakravarty, A.T. Dorsey, M.P.A.
Fisher,
A. Garg, W. Zwerger, Rev. Mod. Phys. 59, (1987) 1.



\bibitem{Guinea} F. Guinea, Phys. Rev. B32 (1985) 4486.

\bibitem{Costi} T.A. Costi, C. Kieffer, Phys. Rev. Lett. 76, (1996) 1683,
cond-mat/9601107. 

\bibitem{Volker} K.  Voelker, Phys. Rev. B58 (1998) 1862, cond-mat/9712080.

\bibitem{Shiba} H. Shiba, Prog. Theor. Phys. 54, (1975) 967.

\bibitem{Shigen} M. Sassetti, U. Weiss,
Phys. Rev. Lett.  65, 2262 (1990).

\bibitem{Grabert} R. Egger, H. Grabert, Phys. Rev. Lett. 75, (1995)
3505;
``Friedel oscillations in Luttinger liquids", in Quantum Transport in
Semiconductors Submicron Structures, edited by B. Kramer, NATO-ASI
Series E (Kluwer, Dordrecht, 1996).

\bibitem{Schmitteckert} P. Schmitteckert, U. Eckern,
Phys. Rev.  B 53, (1996) 15397; cond-mat/9604005.

\bibitem{Fabrizio} M. Fabrizio, A. O. Gogolin, cond-mat/9504011

\bibitem{LLS} A. Leclair, F. Lesage, H. Saleur,  Phys. Rev.
 B 54 (1996) 13597,
cond-mat/9606124.

\bibitem{Chatterjee}, A. Zamolodchikov, R. Chatterjee, hep-th/9311165

\bibitem{CardyLewellen}J. Cardy, D. Lewellen, Phys. Lett. 259B (1991) 274.


\bibitem{LSS} F. Lesage, H. Saleur, S. Skorik, Phys. Rev. Lett. 76,
(1996)
3388, cond-mat/9512087; Nucl. Phys. B 474 (1996) 602,
cond-mat/9603043.


\bibitem{Ghoshal} S. Ghoshal, Int. J. Mod. Phys. A, {\bf 9}, (1994) 4801.

\bibitem{Yurov}V. P. Yurov, A. B. Zamolodchikov, Int.J.Mod.Phys. {\bf
A} 6, (1991) 3419

\bibitem{MussardoCardy} J. Cardy, G. Mussardo, Nucl.Phys. {\bf B}
340, (1990) 387

\bibitem{Affleck} V. Barzykin and I. Affleck, 
cond-mat/9708039

\bibitem{CM} J. Cardy and G. Mussardo, Nucl. Phys. B410 (1983) 451.

\bibitem{BN} J. Balog and M. Niedermaier, Nucl. Phys. B500 (1997) 421. 

\bibitem{Essler} D. Controzzi, F. H. L. Essler and A. M. Tsvelik,
``Optical conductivity of one-dimensional Mott insulators'', cond-mat/0005349

\bibitem{DPTW} P. Dorey, M. Pillin, R. Tateo and G.M.T. Watts, ``One point functions in perturbed boundary conformal field theories'', hep-th/0007077.








\end{thebibliography}
\end{document}